\documentclass[11pt,a4paper]{article}
\usepackage{jheppub}
\usepackage{changepage}
\usepackage{graphicx}
\usepackage{amssymb}
\usepackage{amsmath}
\usepackage{topcapt}
\usepackage{enumitem}
\usepackage{epstopdf}
\usepackage[bottom]{footmisc}
\DeclareMathSymbol{\IR}{\mathbin}{AMSb}{"52}

\preprint{DIAS-STP-14-08\\}
\title{A Quantum Critical Point from Flavours on a Compact Space.}

\author{Veselin G. Filev}

\affiliation{School of Theoretical Physics, Dublin Institute for Advanced Studies\\
10 Burlington Road, Dublin 4, Ireland.}
\emailAdd{vfilev@stp.dias.ie}

\abstract{We analyse a $2+1$ dimensional defect field theory on a two sphere in an external magnetic field. The theory is holographically dual to probe D5-branes in global AdS$_5\times S^5$ background. At any finite magnetic field only the confined phase of the theory is realised. There is a first order quantum phase transition, within the confined phase of theory, ending on a quantum critical point of a second order phase transition. We analyse the condensate and magnetisation of theory and construct its phase diagram. We study the critical exponents near the quantum critical point and find that the second derivatives of the free energy, with respect to the bare mass and the magnetic field, diverge with a critical exponent of $-2/3$. Next, we analyse the meson spectrum of the theory and identify a massless mode at the critical point signalling a diverging correlation length of the quantum fluctuations. We find that the derivative of the meson mass with respect to the bare mass also diverges with a critical exponent of $-2/3$. Finally, our studies of the magnetisation uncover a persistent diamagnetic response similar to that in mesoscopic systems, such as quantum dots and nano tubes.}


\subheader{}

\begin{document}
\maketitle
\flushbottom
\section{Introduction}
Quantum phase transitions are phase transitions which take place at zero temperature and are thus driven by quantum as opposed to thermal fluctuations. Such phase transitions are realised by tuning a non-temperature control parameter such as pressure, chemical potential or magnetic field. Of particular interest are second order quantum phase transitions which, similarly to their thermal analogue, have a quantum critical point where the quantum fluctuations driving the transition diverge and become scale invariant. The properties of these systems near the phase transition are characterised by critical exponents, which are independent on the microscopic details of the systems -- a property called Universality. At the critical point fluctuations of all size scales become important demanding a scale invariant description, naturally provided by conformal field theories. Universality can then be seen as a by-product of the fact that there are relatively few scale invariant theories. 

Although quantum phase transitions take place at zero temperature, at finite temperature quantum fluctuations compete with thermal fluctuations and the existence of a ``quantum critical'' regime is expected, bearing remnants of the quantum critical point even at finite temperature. Experimentally the quantum critical regime becomes manifest in an exotic physical behaviour often signalling the existence of a novel phase. 

Quantum critical points are an active area of research in condensed matter community, where they appear in the phase diagram of high temperature superconductors~\cite{Sachdev}. Clearly the existence of a quantum critical regime accessible at finite temperature is crucial for such applications, however away from the critical point one needs more than just an effective field theory and the analysis often relies on physical arguments and approximations rather than rigorous calculations. It is therefore, of a particular importance to come up with clean theoretical models, which not only allow full analytic (or quasi numerical) control, but are complex enough to release the vicinity of a quantum critical point. The holographic quantum critical point that we propose in this paper is an example of such a model and is potentially applicable to the qualitative description of the quantum critical regions in realistic condensed matter systems such as high temperature superconductors. 

The original AdS/CFT correspondence establishes a duality between an ${\cal N}=4$ super Yang-Mills theory (SYM) in $3+1$ dimensions, which is a conformal field theory, and superstring theory on an AdS$_5\times S^5$ space-time. There is no known quantum critical point described by ${\cal N}=4$ SYM. In fact, although a vast amount of novel supergravity backgrounds dual to conformal field theories have been constructed, usually their construction is not associated with the critical regime of a particular theory. As a result relatively few holographic quantum critical points have been realised, usually in the context of holographic flavour dynamics. 

A system with continuous quantum phase transition has been constructed in ref.~\cite{Karch:2009ph}, however this construction involved a third order phase transition. A second order quantum phase transition has been realised in ref. \cite{Karch:2007br}, where the holographic gauge theory dual to the D3/D7 intersection at zero temperature and finite chemical potential has been explored. A quantum critical point has also been constructed in refs \cite{Jensen:2010vd,Evans:2010iy}, where the holographic gauge theory dual to the D3/D5 intersection at finite magnetic field and chemical potential has been analysed. In this paper we porpoise a novel quantum critical point realised in a 1+2 dimensional defect field theory living on a maximal two sphere and subjected to an external magnetic field. 

Our holographic set up involves probe D5-branes in a global AdS$_5\times S^5$ space-time. The D5-branes are extended along the radial direction and wrap a maximal two sphere inside the AdS$_5$ part of the geometry. The D5-branes also wrap a two sphere inside the $S^5$ part of the geometry. The corresponding gauge theory is an ${\cal N}=4$ SYM theory on a three sphere coupled to an ${\cal N}=2$ hypermultiplet living on a maximal two sphere. The system is supersymmetric only at vanishing bare mass, when the D5-brane wraps a maximal two sphere inside the $S^5$. At finite bare mass supersymmetry is broken and the theory develops a fundamental condensate. 

The properties of the theory at zero magnetic field can be further analysed by considering a pair of external probe quarks. A key observation is that at strong coupling the screening length is inversely proportional to the bare mass \cite{Karch:2009ph}. This suggests that for large bare masses the screening length is small compared to the radius of the two sphere and the theory is in a confined phase (separating the quarks would result in a pair production). For small bare  masses the screening length is larger than the radius of the sphere and a pair of quarks can be separated without pair production -- the theory is in a deconfined  phase. The quantum phase transition between the two regimes is of a first order \cite{Karch:2009ph} and is triggered by the Casimir energy of the theory.  In the holographic set up the phase transition is realised
 as a topology change transition for D5-brane embedding. The confined phase of the theory corresponds to embeddings, which close above the origin of AdS$_5$ by having the $S^2$ inside the $S^5$ part of the geometry shrink to zero size, referred to as ``Minkowski'' embeddings, while the deconfined phase of the theory corresponds to embeddings which reach all the way to the origin of AdS$_5$ and have the $S^2$ inside the AdS$_5$ part of the geometry shrinking, we will refer to them as the ``Ball'' embeddings. 
\changepage{1em}{}{}{}{}{}{}{}{-1em}

At first glance subjecting the theory to an external magnetic field seems like an innocent exercise. Indeed, the second betti number of the two sphere is non-zero and turning on a magnetic field on the two sphere does not require additional electric current. In the holographic set up this can be achieved by either introducing a pure gauge $B$-field or by fixing the $U(1)$ gauge field of the D5-brane. However, although such solutions can be constructed relatively easy, for ball embeddings one encounters a problem. At the origin of AdS$_5$ the radius of the two sphere shrinks to zero size, but the magnetic flux throughout the two sphere remains finite, as a result the norm of the $U(1)$ field strength diverges at the origin, signalling the presence of a magnetic monopole. One can show that a magnetic monopole on the world volume of the D5-brane would violate charge conservation of the RR-flux unless some number of D3--branes are attached at the position of the monopole. On the other hand, doing so would deform the D5-brane embedding due to the tension of the D3--branes. 

In general constructing a balanced D3-D5-brane configuration can be a challenging task, but in our case a small miracle happens: the ``Ball''  D5-brane embeddings are mimicked by Minkowski embeddings, which close at very small but finite distance above the origin of AdS thus avoiding divergent norm of the $U(1)$ field strength and the need of a magnetic monopole. Furthermore, these embeddings fold along the $S^5$ part of the geometry, while still close to the origin, developing a $\IR\times B_{3}\times S^2$ throat with very small radius of the $S^2$, mimicking the D3--branes needed to source the magnetic monopole in the unbalanced D3-D5- configuration (see figure~\ref{fig:fig3}). 

One can show that Minkowski embeddings cover the whole parametric space, completely replacing the ```Ball'' embeddings, this picture persist even for an infinitesimal magnetic field. The first order confinement/deconfiment phase transition also has its analogue within the confined phase described by the Minkowski embeddings.\footnote{This situation is the magnetic analogue of the finite density study of ref.~\cite{Kobayashi:2006sb}, where the radial electric field of the probe D7-brane demands the presence of an electric monopole for Minkowski embeddings. The fundamental strings required to source the monopole are realised as spikes of black hole embeddings mimicking Minkowski embeddings.} However, unlike the phase transition at vanishing magnetic field, there is no associated topology change for the D5-brane embeddings (only an approximate one between embeddings with and without an $\IR\times B_{3}\times S^2$ throat) as a result we find that at sufficiently strong magnetic field the first order phase transition ends on a critical point of a second order phase transition. Increasing further the magnetic field leads to the replacement of the phase transition by a smooth cross over. Focusing on the properties of the theory near the quantum critical point we find that all second derivatives of the free energy diverge as functions of the bare mass and the external magnetic field with critical exponent $-2/3$. In addition, our studies of the meson spectrum show that at the critical point there is a massless mode which we associate to the divergent correlation length of the quantum fluctuations. The derivative of the mass of this mode with respect to the bare mass also diverges at the phase transition with a critical exponent $-2/3$.

Another interesting property of the theory is its magnetic response. Our studies of the magnetisation and magnetic susceptibility show that the theory is a diamagnetic, which is not unusual for holographic gauge theories (see for example ref.~\cite{Filev:2013vka}). However, our system has peculiar properties in the limit of vanishing magnetic field. The analogue of the deconfined phase has a persistent diamagnetic response independent on the magnitude of the magnetic field. This behaviour is similar to the persistent diamagnetic current observed in realistic mesoscopic systems such as nano tubes and quantum dots \cite{Zelyak}. The fact that this phase is realised for small radius of the two sphere, when the size of the system is small and finite size quantum effects (such as the existence of Casimir energy) are important, makes the comparison to mesoscopic systems plausible. Furthermore, the fact that the observed effect is in the strongly coupled regime of the system suggests potential applications in the description of strongly coupled condensed matter systems, exhibiting persistent diamagnetic response. 

Finally, for any finite value of the magnetic field the sable phase of the theory at vanishing bare mass has a negative condensate. This suggests that the global symmetry corresponding to rotations in the transverse space to the D5-branes is spontaneously broken. In this way our system realises the effect of magnetic catalysis of chiral symmetry breaking confirming the universal nature of this phenomena \cite{mag-cat}. \footnote{For recent holographic studies of this phenomena in $1+2$ dimensional systems see refs.~ \cite{Evans:2013jma,Kristjansen:2013hma, Filev:2014bna}} 

The paper is organised as follows: In section \ref{sec2} we review the properties of the theory at zero magnetic field studied in refs. \cite{Karch:2009ph}, \cite{Erdmenger:2010zm}. In section \ref{sec3} we explore the properties of the theory at finite magnetic field. We show that at any finite magnetic field only Minkowski embeddings are physical and only the confined phase of theory is realised. We study the fundamental condensate of the theory as a function of the bare mass and show that for sufficiently small magnetic fields there is a first order phase transition within the confined phase, which ends on a critical point of a second order phase transition. We also calculate the magnetisation of the theory and its magnetic susceptibility. Finally, we study the critical behaviour of the theory near the quantum critical point and calculate various critical exponents. In section \ref{sec4} we analyse the meson spectrum of the theory. We obtain the general quadratic action for all four scalar and six vector meson modes. We then focus on the spectrum of one of the scalar modes, which becomes massless at the quantum critical point, corresponding to the divergent correlation length of the quantum fluctuations. Finally, we focus on the ground state of the spectrum near the phase transition and calculate the corresponding critical exponent. Section \ref{sec5} contains our conclusion.
%
\vspace{-.35em}
\section{The theory at zero magnetic field}\label{sec2}
The supergravity dual of ${\cal N}=4$ SYM theory on a three sphere is an AdS$_5\times S^5$ space-time in global coordinates with a metric given by:
\begin{equation}
ds^2=-\left(1+\frac{r^2}{R^2}\right)dt^2+r^2\,d\Omega_3^2+\frac{dr^2}{1+\frac{r^2}{R^2}}+R^2(d\theta^2+\cos^2\theta\, d\Omega_2^2+\sin^2\theta\, d\tilde\Omega_2^2)\ .
\end{equation}
\changepage{-1em}{}{}{}{}{}{}{}{1em}

To introduce fundamental flavours in the quenched approximation to the dual field theory we will consider a probe D5-brane extended along the radial coordinate $r$ and wrapping two spheres in both the $S^5$ and the AdS$_5$ parts of the geometry. More explicitly if we parameterise the $S^3\subset $ AdS$_5$ and one of the $S^2\subset S^5$ by:
\begin{eqnarray}
&&d\Omega_3^2=d\alpha^2+\sin^2\alpha\, (d\beta^2+\sin^2\beta \,d\gamma^2)\ , \\
&&d\Omega_2^2=d\tilde\alpha^2+\sin^2\tilde\alpha \,d\tilde\gamma^2\nonumber\ ,
\end{eqnarray}
then the D5-brane probe is extended along the $t,\alpha\,,\gamma\, ,r\,,\tilde\alpha\,$ and $\tilde\gamma$ directions. Note that the D5--brane respects the $\mathbb Z^2$ symmetry of the dial defect field theory by wrapping completely the boundary $S^2$ in the AdS$_5$ part of the geometry. One can check that placing the D5-brane at $\beta=\pi/2$ and  allowing the D5-brane to have a non-trivial profile along $\theta$ is consistent with the equations of motion. The most symmetric ansatz (which is also consistent with the general equations of motion) is to consider $\theta =\theta(r)$. The corresponding field theory is coupled to an ${\cal N}=2$ fundamental hypermultiplet living on the two sphere.  The theory is at most ${\cal N}=2$ supersymmetric, and one can show that a non-trivial profile $\theta(r)\neq 0$ completely breaks supersymmetry. When supersymmetry is broken the theory develops a fundamental condensate, with non-trivial dependence on the bare mass of the theory. Furthermore, the theory has a first order phase transition corresponding to a topology change transition in the holographic set up. Indeed, generally the possible embeddings of the D5-brane split into two classes - embeddings which close at some radial distance above the origin of AdS$_5$ by having the $S^2$ inside the $S^5$ part of the geometry shrink to zero size (Minkowski embeddings), and embeddings which reach all the way to the origin of AdS$_5$ and close there, which we call ``Ball'' embeddings. This construction has been investigated in ref.~\cite{Karch:2009ph}, where it has been shown that the phase transition is of a first order. It has also been argued that this is a confinement/deconfinement phase transition driven by the Casimir energy of the theory on $S^2$. The properties of this theory have been further investigated in ref. \cite{Erdmenger:2010zm}, where the meson spectrum of the theory has been studied in details. 

In what follows, we will briefly review the properties of the theory at zero temperature and external magnetic field. For illustrative purposes it is instructive to introduce a new radial coordinate $u(r)$ given by:\footnote{Note that in these coordinates the origin of AdS$_5$ is at $u =R/2$. }
\begin{equation}\label{u_of_r}
u=\frac{1}{2}(r+\sqrt{R^2+r^2})\ ,
\end{equation}
which enables us to write part of the geometry as a conformal $\IR^6$:
\begin{equation}
ds^2=-\frac{u^2}{R^2}\left(1+\frac{R^2}{4u^2}  \right)^2\,dt^2+u^2\,\left(1-\frac{R^2}{4u^2}  \right)^2\,d\Omega_3^2+\frac{R^2}{u^2}\,(du^2+u^2\,d\Omega_5^2)\ .
\end{equation}
The conformally $\IR^6$ part of the geometry can be split into a conformally $\IR^3$ part wrapped by the D5-brane and a conformally $\IR^3$ part transverse to the D5-brane:
\begin{equation}
du^2+u^2\,d\Omega_5^2=d\rho^2+\rho^2\,d\Omega_2^2+d l^2+l^2\,d\tilde\Omega_2^2\ ,
\end{equation}
where $\rho= u\,\cos\theta$ and $l =u\sin\theta$. The possible D5-brane embeddings can then be visualised in the $\rho$ versus $l$ plane. While we will eventually present our results in these coordinates, technically it is more convenient to work in the original ($r$,$\theta$) coordinates. For the radial part of the corresponding DBI action (after integrating over the $\tilde S^2\subset S^5$ and the $S^2\subset $ AdS$_5$) one obtains:
\begin{equation}\label{DBI-zero-B}
{\cal L} \propto \,r^2\cos^2\theta\sqrt{1+(r^2+R^2)\,\theta'(r)^2}\ .
\end{equation}
The solution to the equation of motion for $\theta(r)$ has the following asymptotic behaviour at large $r$:
\begin{equation}
\theta(r)=\frac{m}{r}+\frac{c}{r^2}+\dots
\end{equation}
and according to the AdS/CFT dictionary, the parameter $m$ is related to the bare mass of the fundamental multiplet $m_q$ via $m_q =m/(2\pi\alpha')$, while the  parameter $c$ is proportional to the fundamental condensate of the theory $\langle\bar\psi\,\psi\rangle \propto -c$. Note that to arrive at the last result one has to introduce appropriate counter terms \cite{Karch:2009ph} (see also \cite{Karch:2005ms}). However, with this definition one obtains a constant non-zero value of the condensate at infinite bare mass, suggesting that an additional counter term is needed.

\begin{figure}[t] 
   \centering
   \includegraphics[width=4.5in]{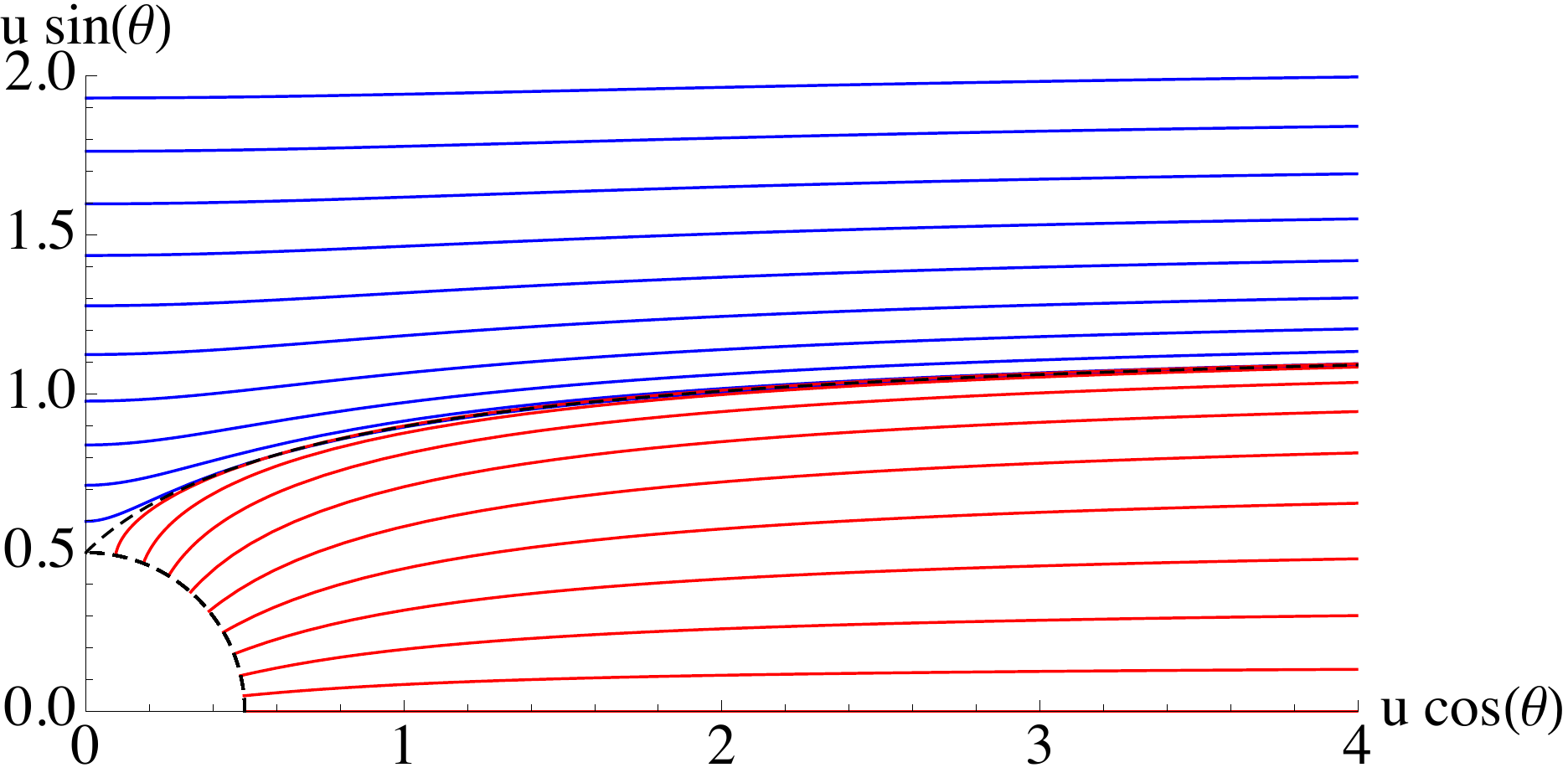} 
   \vspace{-.5em}
   \caption{\small A plot of the possible D5-brane embeddings. Red curves represent ``Ball'' embeddings closing at the origin, while blue curves correspond to ``Minkowski'' embeddings closing above the origin. The two classes are separated by a critical embedding (the black dashed curve) having both the $\tilde S^2\subset S^5$ and $S^2\subset $ AdS$_5$ shrinking at the origin. The plot is in units $R=1$.}
   \label{fig:fig1}
\end{figure}

\begin{figure}[htbp] 
   \centering
   \includegraphics[width=4.5in]{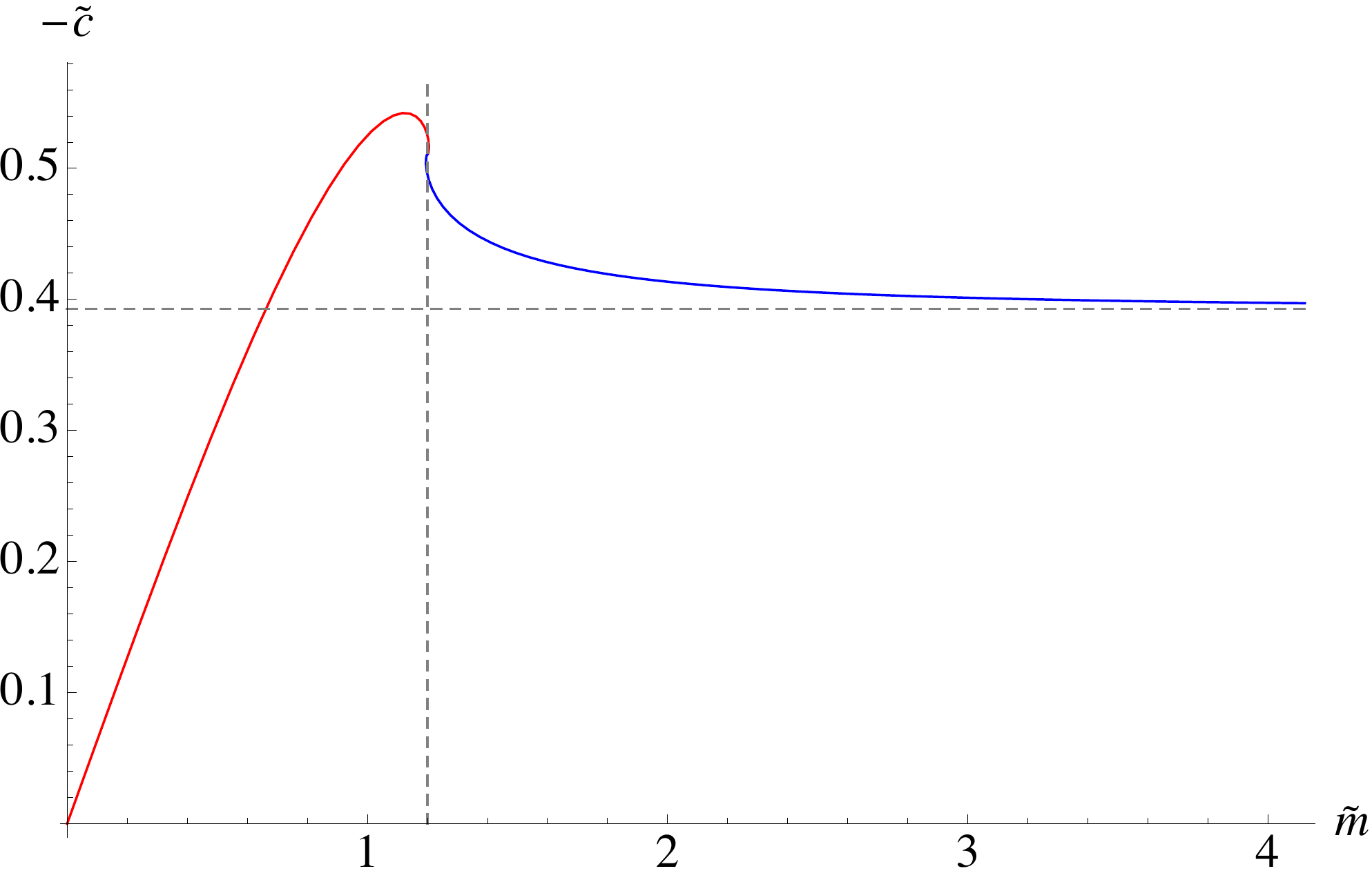} 
   \caption{\small A plot of the condensate parameter $\tilde c =c/R^2$ versus the bare mass parameter $\tilde m =m/R$. The blue curve represents the confined phase (Minkowski embeddings), while the red curve represents the deconfined phase (``Ball'' embeddings). At $\tilde m\approx 1.20$ the theory undergoes a first order phase transition as evident by the multivalued dependence of $-\tilde c$ near the critical state.   }
      \label{fig:fig2}
\end{figure}

In general the equation of motion for $\theta(r)$ derived from (\ref{DBI-zero-B}) has to be solved numerically. One can check that for both Minkowski and ``Ball'' embeddings, the equation of motion has predetermined boundary conditions at the closing point, in the sense that the solution depends only on one parameter. This enables us to obtain a perturbative analytic solution, which we feed into a numerical shooting technique. A set of possible D5-brane embeddings is presented in figure \ref{fig:fig1}. The red curves represent Ball embeddings reaching the origin (the black segment), while the blue curves represent Minkowski embeddings. One can see that the two classes are separated by a critical embedding (the dashed black curve), which has both the $\tilde S^2\subset S^5$ and the $S^2\subset $ AdS$_5$ shrinking at the origin. 
By extracting the coefficients $m$ and $c$ from the asymptotic behaviour of the embeddings at large radial distances $r$, one can generate the plot of the condensate versus bare mass presented in figure~\ref{fig:fig2}, where we have introduced the dimensionless parameters $\tilde m=m/R$ and $\tilde c =c/R^2$. Note, that the physical meaning of the parameter $\tilde m$ follows from the relation $\tilde m =  m\,R /R^2\propto m_q R_2 /\sqrt{\lambda}$, where $R_2$ is the radius of the defect $S^2$  (which is equal to radius of the field theory $S^3$) and $\lambda$ is the t'Hooft coupling. At large $\tilde m$ the radius of the two sphere is large (at fixed bare mass), the Casimir energy is small and the theory is in a confined phase (see refs. \cite{Karch:2009ph},\cite{Karch:2006bv}). As $\tilde m$ decreases the Casimir energy increases and at $\tilde m\approx 1.20$ there is a first order confinement/deconfinement phase transition triggered by the Casimir energy. One can also see from the plot that at large $\tilde m$ the condensate approaches a constant value. In fact one can show analytically that the parameter $\tilde c$ approaches $\pi/8$. Since large $\tilde m$ corresponds to large bare masses this is a surprising result. One would expect that as the flavours become infinitely massive they will decouple from the theory and hence the fundamental condensate will vanish. This suggests that a finite counter term (that was not prescribed in refs. \cite{Karch:2009ph},\cite{Karch:2005ms}) is needed. 

Given that the condensate approaches a constant value at large bare mass, and hence the free energy grows linearly with the bare mass, the appropriate counter term should be linear in the supergravity field $\theta$. In fact, in order to preserve the $\theta\to-\theta$ symmetry of the Lagrangian (\ref{DBI-zero-B}), it should be proportional to $|\theta|$. Indeed, one can show that a counter term proportional to $\sqrt{-\gamma}\, R_{\gamma}\,|\theta|$, where $\gamma$ is the induced metric on the asymptotic boundary spanned by the field theory directions of the defect, and $R_{\gamma}$ is the corresponding scalar curvature, produces a term linear in $|m|. $\footnote{Note that naively this choice of a counter term breaks the scaling symmetry of the boundary field theory, However, the flavoured gauge theory is conformal only for vanishing bare mass, when $\theta\equiv 0$ and the counter term vanishes. One can show that for $\theta\neq 0$ both conformal symmetry and supersymmetry are explicitly broken and hence the proposed counter term does not break any further symmetries.}  However, this counter term induces a discontinuity of the condensate of the theory because $\partial_m |m| =\text{sign}(m)$. Since we are mainly interested in the properties of the theory near the phase transition, we will take the same approach as in ref. \cite{Karch:2009ph} and use the present definition of the condensate (being proportional to $-c$).

\section{The theory at finite magnetic field}\label{sec3}
\subsection{General set up}
We are interested in the properties of the theory at finite magnetic field, which (as advertised in the introduction) realises a quantum critical point. Because the second betti number of the two sphere is non-vanishing a magnetic field can be turned on without the introduction of external electric currents. A natural choice in this case is to consider a magnetic field proportional to the volume form of the two-sphere. In order to turn on such a magnetic field, we turn on a pure gauge B-field:
\begin{equation}\label{B2}
B_{(2)}= H R^2 \sin\alpha\,d\alpha\wedge d\gamma\ .
\end{equation}
One can check that $B_{(2)}=d\Lambda$ for $\Lambda = -HR^2 \cos\alpha \,d\gamma$. Thus equivalently, we could introduce magnetic field by turning on the $A_{\gamma}$ component of the U(1) gauge field. In both cases one should check\footnote{Note that as pointed out in ref. \cite{Chunlen:2014zpa} the analogous ansatz in the case of D7-brane probe considered in ref. \cite{Filev:2012ch} does not satisfy the equations of motion for the gauge field. The physical reason for that is that the second betti number of the three sphere is zero and one has to introduce external currents to support such a magnetic field. The study considered in ref.\cite{Filev:2012ch} has been reinterpreted as a bottom up study, where one introduces an external $F_7$ flux (serving as external currents) to support the magnetic field on the D7-brane.} that the equations of motion for the gauge field are satisfied. The DBI action of the D5-brane is then given by:
\begin{equation}\label{D5-DBI}
S_{DBI}=-\frac{\mu_5}{g_s}\int d^6\xi \,e^{-\Phi} \sqrt{-\text{det} (P[G+B]+2\pi\alpha'F)}\ ,
\end{equation}
defining $E=P[G+B]+2\pi\alpha' F$ for the equation of motion for the gauge field we obtain:
\begin{equation}\label{A}
\partial_{\nu}(\sqrt{-E}\,(E^{\mu\nu}-E^{\nu\mu}))=0.
\end{equation}
One can easily check that for the ansatz $\theta=\theta(r)$, $A_\mu =0$ and $B_{(2)}$ given by equation (\ref{B2}), the equation of motion for the gauge field (\ref{A}) is satisfied. The equation of motion for $\theta(r)$ is derived by varying the DBI action (\ref{D5-DBI}):
\begin{equation}\label{lag-H}
S_{DBI}=-\frac{\mu_5\,(2\pi R)^2}{g_s}\int dt\,dr\sqrt{r^4+R^4H^2}\,\cos^2\theta(r)\sqrt{1+(R^2+r^2)\theta'(r)^2}\ .
\end{equation}
Naively, one expects that for moderate magnetic field the behaviour of the system should be similar to the zero magnetic field case. One expects that the embeddings would again split into ``Ball'' and ``Minkowski'' classes and there would be a first order phase transition between the two, at some critical value of the parameter $\tilde m$. Furthermore, because of the universal nature of the effect of magnetic catalysis, one would expect that at strong magnetic fields the phase transition disappears and only the confined phase, with chiral symmetry breaking vacuum at vanishing bare mass, is stable. However, as we are going to see, something much more intricate happens and only part of our naive expectations are met. 

\subsection{Branes and magnetic monopoles}
Let us calculate the norm of the B-field $|B|^2=\frac{1}{2}B_{\mu\nu}B^{\mu\nu}$ , we obtain: $|B|^2=R^4H^2/r^4$. Therefore, for ``Ball'' embeddings the norm of the B-field diverges at the origin of AdS$_5$. This simply reflects the fact that for our ansatz the magnetic flux through the $S^2$ of the defect remains constant, and since at the origin of AdS the two-sphere shrinks to zero size, there should be a magnetic monopole\footnote{Note that by magnetic monopole we mean a charged object S-dual to an electric monopole.} sitting there. The world volume of the D5-brane is six dimensional and in six dimensions a magnetic monopole is a two dimensional object. In our case the monopole should be wrapping the $S^2\subset S^5$, which remains finite at the origin. It has been known for quite a while that a Dp-2 brane ending on a Dp-brane sources a magnetic monopole in the world volume of the Dp-brane \cite{Strominger:1995ac}. We can easily see this. The Wess-Zummino term of the D5-brane contains a coupling of the form:
\begin{equation}
\mu_5\int B_{(2)}\wedge C_{(4)}\ , 
\end{equation}
where $C_{(4)}$ is a four form Ramond-Ramond potential. Now gauge invariance of the Wess-Zummino action under the transformation $C_{(4)} \to C_{(4)} +d\Lambda_{(3)}$ suggests that the term:
\begin{equation}\label{gauge tr}
\mu_5\int B_{(2)}\wedge d\Lambda_{(3)} = -\mu_5\int dB_{(2)}\wedge\Lambda_{(3)} =-\mu_5\,(2\pi R^2)\,H\int_{\Sigma}\Lambda_{(3)}
\end{equation}
should vanish, where we have used that $dB_{2} =HR^2\,{\delta(r)}\,\sin\alpha \,dr\wedge d\alpha\wedge d\gamma $ and $\Sigma$ is spanned by the time and $S^2\subset S^5$ directions at the origin of AdS. Therefore, we see that the magnetic monopole induces a three dimensional ``boundary'' of the D5-brane and to preserve gauge invariance we have to attach some number ($N_3$) of D3-branes to the D5-brane. The variation of the Wess-Zummino action of the D3-branes with respect to the gauge transformation of $C_{(4)}$ would result in a term $-N_3\,\mu_3\int_{\Sigma} \Lambda_{(3)}$, which can be used to cancel the term in equation (\ref{gauge tr}). For the number of D3-branes required to source the magnetic monopole we obtain:
\begin{equation}\label{N3}
N_3 =\frac{\mu_5}{\mu_3}\,(2\pi R^2)\,H\ .
\end{equation}
\begin{figure}[t] 
   \centering
   \includegraphics[width=4.3in]{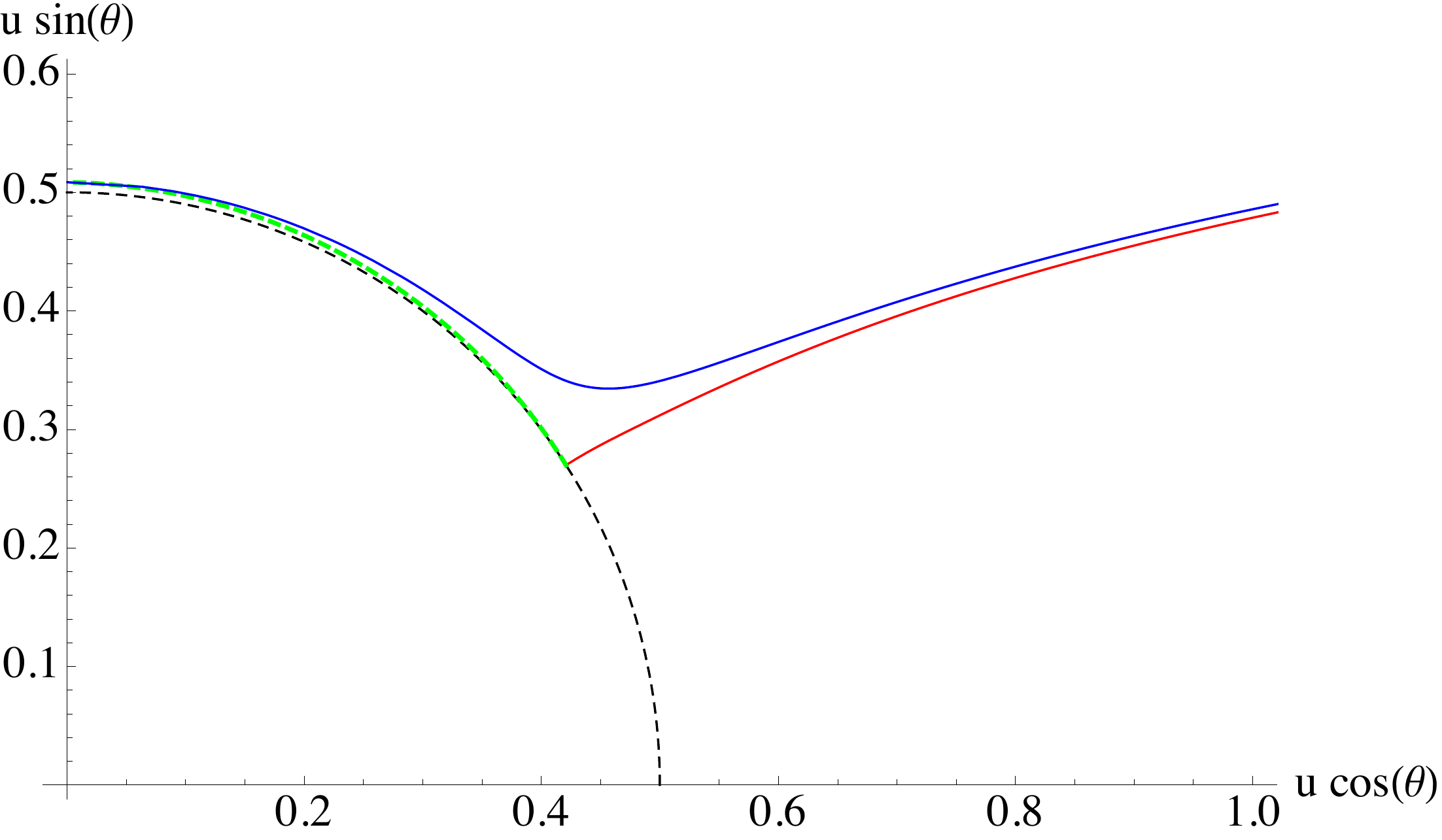} 
   \caption{\small  A plot of a typical ``Ball'' embedding  (red curve) and a D3-brane attached to it (green dashed segment). The D3-brane ends at the D5-brane at $(\theta =0.57, r=0)$ and closes at $\theta=\pi/2$ for some $r_0$, thus having a topology $\IR\times B_3$. The tension of the D3-brane deforms the D5-brane. The stable configuration is a Minkowski D5-brane embedding which realises the D3-brane as a throat of topology $\IR\times B_3\times S^2$, with a tiny radius of the $S^2$. The $S^2\subset$ AdS$_5$ never shrinks completely, avoiding the need of a localised magnetic monopole.}
      \label{fig:fig3}
\vspace{-.3em}
\end{figure}

To visualise the D3-branes, that we need to attach to the D5-brane, let us plot a typical ``Ball'' D5-brane embedding. The resulting plot (for $H=0.01$) is presented in figure~\ref{fig:fig3}. The solid red curve represents a ``Ball'' D5-brane embedding ending at the origin at $\theta=\theta_0=0.57$. The green dashed segment represent a D3-brane sitting near the origin of AdS$_5$, wrapping an $S^2\subset S^5$ and extended along $\theta$ and $r$, The D3-brane ends on the D5-brane at $(\theta=\theta_0,r=0)$, and closes smoothly at $(\theta =\pi/2,r=r_0)$ thus having a topology  $\IR\times B_3$. Of course, this configuration is not stable. The tension of the D3-brane will pull the D5-brane and the configuration will change. Remarkably, the stable configuration is found among the Minkowski class of embeddings! The blue curve represents a Minkowski embedding mimicking the D3-D5 configuration. The D3-brane is realised by the Minkowski embedding as a throat of topology $\IR\times B_3\times S^2$, where the $S^2\subset$ AdS$_5$ has a very small radius along the segment near the origin. Note that because Minkowski embeddings close above the origin, the $S^2\subset$ AdS$_5$  never shrinks to zero size, avoiding the need of a localised magnetic monopole.\footnote{Note that this situation is a magnetic analogue of the finite density study of ref. \cite{Kobayashi:2006sb}, where the radial electric field of the probe D7-brane requires the presence of an electric monopole for Mikowski embedding, the fundamental strings required to source the monopole are realised as spikes of black hole embeddings mimicking Minkowski embeddings.} 

\changepage{1.1em}{}{}{}{}{}{}{}{-1.1em}
In fact, we can present further evidence that the throat of the D5-brane realises D3-branes. Using that it is located in a region $r_0<r<r_0+\delta r\ll1$, for the part of the DBI action corresponding to the throat (to leading order) we obtain:
\vspace{-.6em}
\begin{align}\nonumber
\delta S_{DBI}&=-\frac{\mu_5\,(2\pi R)^2 }{g_s}R^2H\int dt\int_{r_0 }^{r_0+\delta r}dr\,\cos^2\theta(r)\sqrt{1+(R^2+r^2)\theta'(r)^2}=\\ 
&=-N_3\frac{\mu_3}{g_s}\int dt \int_{r_0}^{r_0+\delta r}dr d\tilde\alpha d\tilde\gamma \sqrt{-G_{tt}\,G_{\tilde\alpha\tilde\alpha}\,G_{\tilde\gamma\tilde\gamma}}\sqrt{G_{rr}+G_{\theta\theta}\theta'(r)^2}\ , \label{DBI-D3}
\end{align}
\vspace{-1em}
where:
\vspace{-.1em}
\begin{align}\nonumber
&G_{tt}=-\left(1+\frac{r^2}{R^2}\right);~~G_{\tilde\alpha\tilde\alpha}=R^2\cos^2\theta;~~G_{\tilde\gamma\tilde\gamma}=\sin^2\tilde\alpha\,G_{\tilde\alpha\tilde\alpha};\\
&G_{rr}=\left(1+\frac{r^2}{R^2}\right)^{-1};~~G_{\theta\theta}=R^2 \ ,
\end{align}
\changepage{-1.1em}{}{}{}{}{}{}{}{1.1em}and we have used equation (\ref{N3}). Note that the integrant in the last equation in (\ref{DBI-D3}) is~the volume form associated to the induced metric on a D3-brane wrapping the $S^2\subset S^5$ and extended along $\theta$, therefore the total expression is precisely the DBI action of a stack of $N_3$ D3-branes wrapping the $S^2\subset S^5$ and extended along $\theta$ and $r$ for $r_0<r<r_0+\delta r$. In the next subsection we will use this set up to study the fundamental condensate of the theory.
\subsection{Fundamental condensate}
Our numerical studies confirm that Minkowski embeddings cover the entire parametric space and are the thermodynamically stable phase for any arbitrary small non-zero magnetic field. What is more, we find that the first order confinement/deconfinement phase transition, at vanishing magnetic field, has an analogue and there is a first order phase transition, within the confined phase, ending on a critical point of a second order quantum phase transition. 
\begin{figure}[t] 
   \centering
   \includegraphics[width=2.55in]{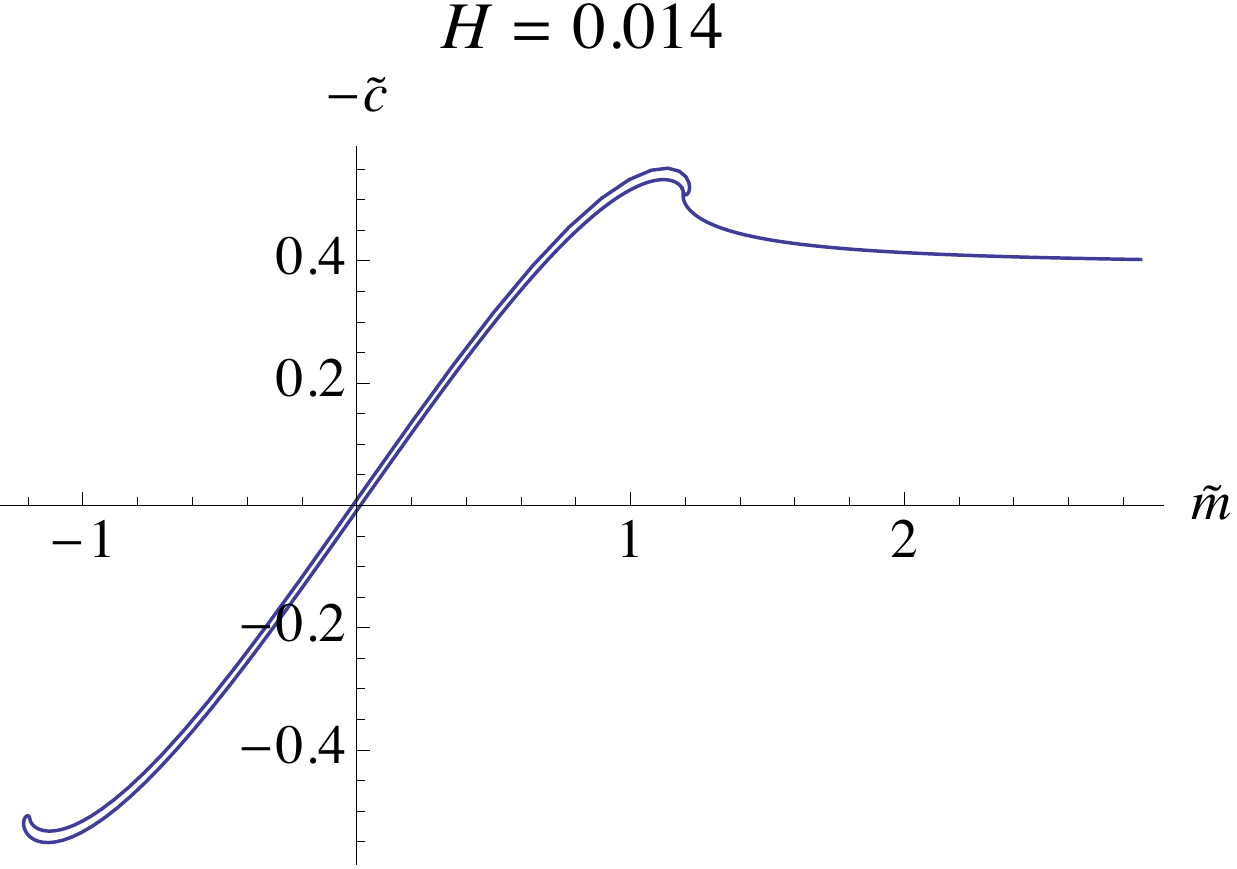} 
  \includegraphics[width=2.55in]{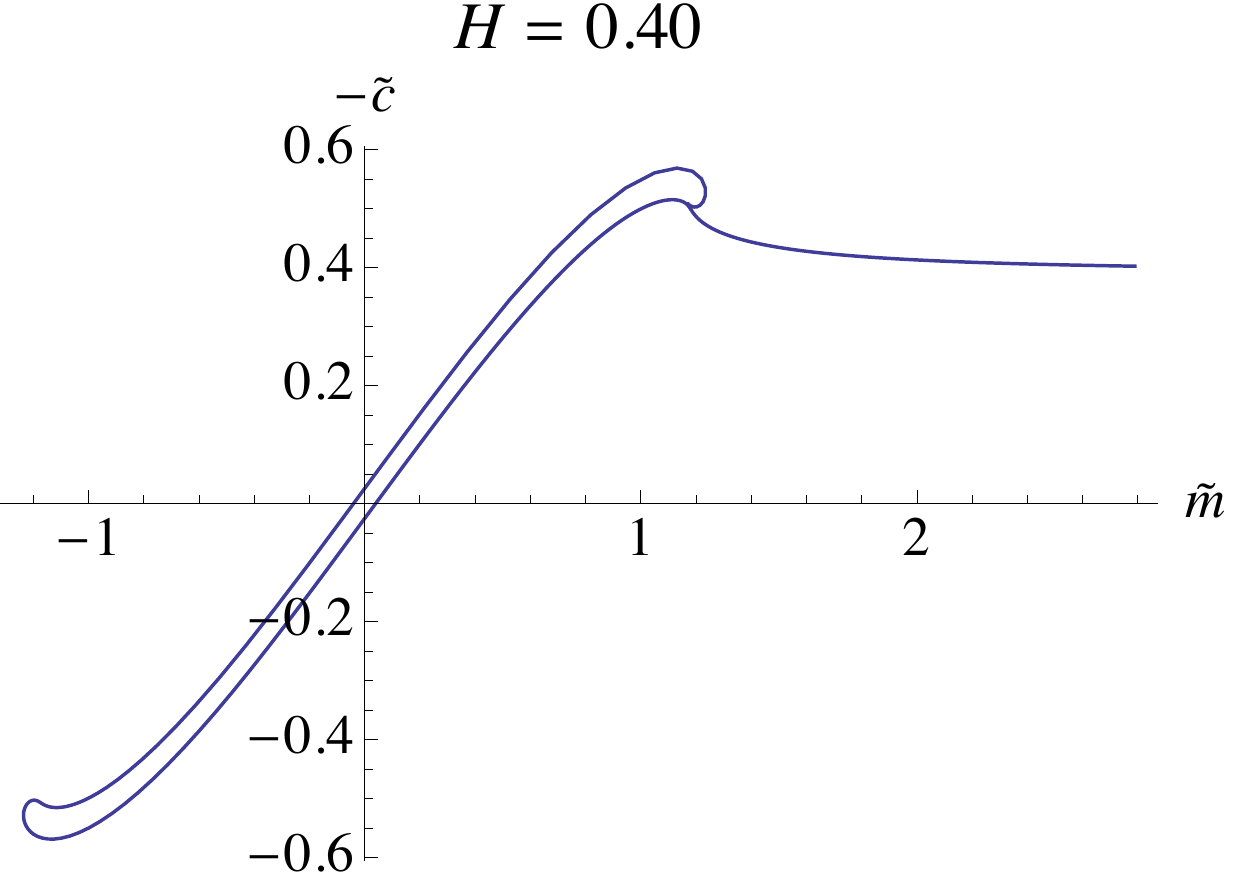} 
  \includegraphics[width=2.55in]{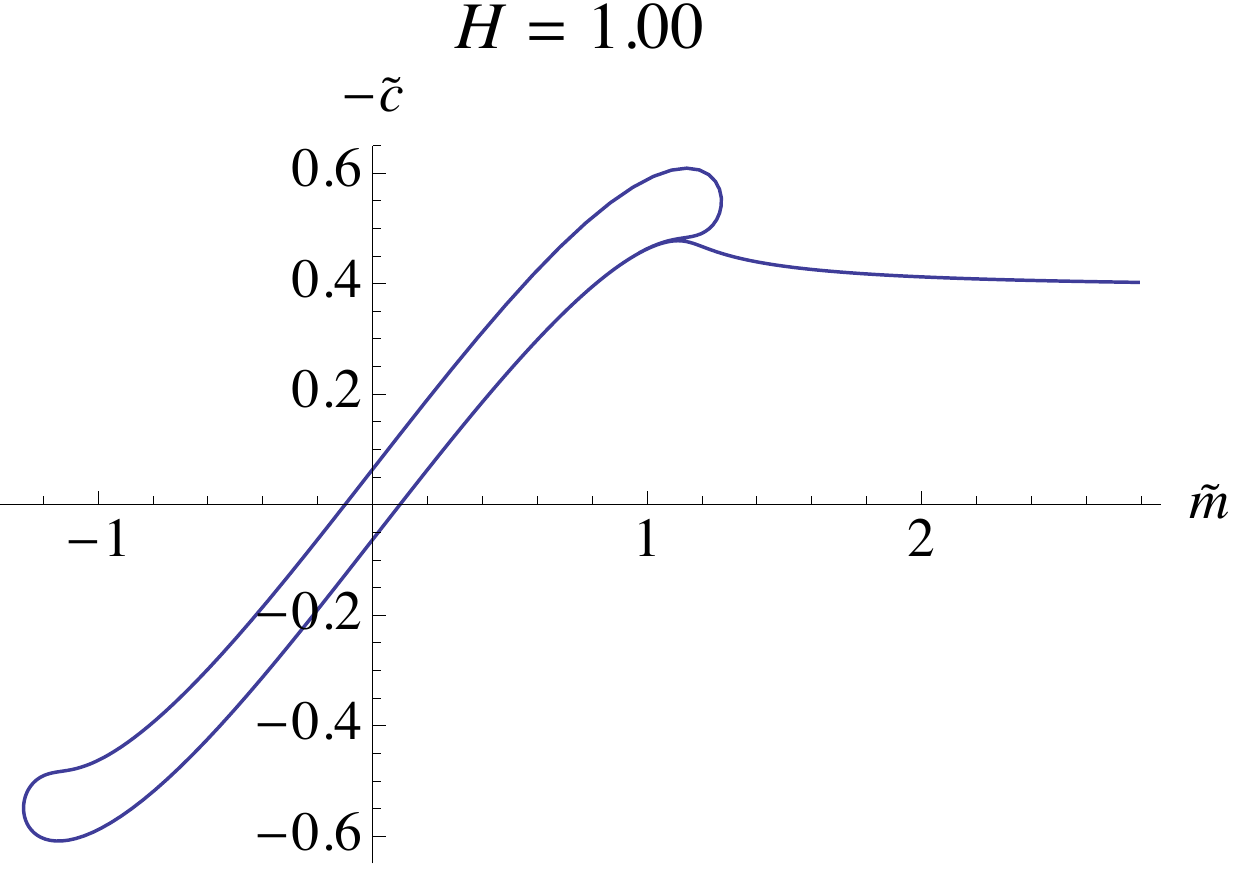}
  \includegraphics[width=2.55in]{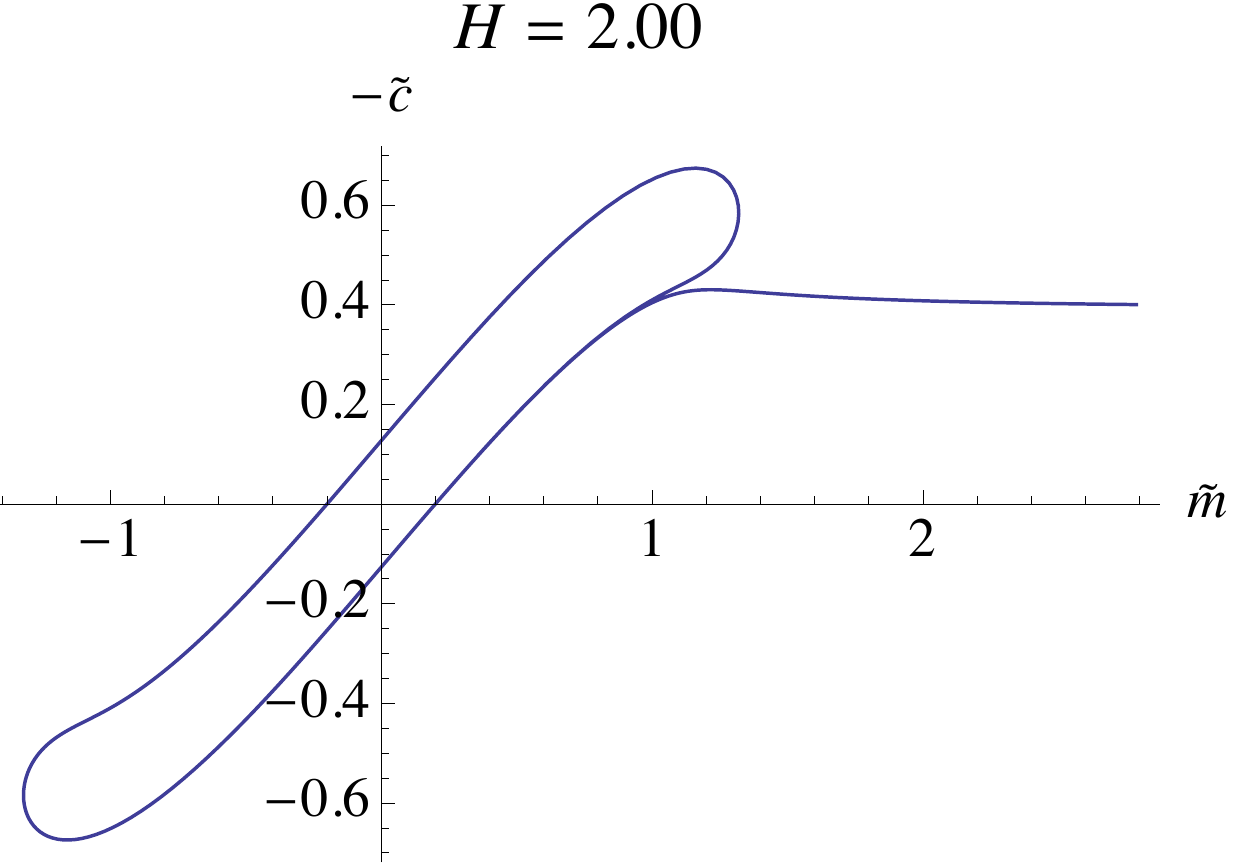}
   \caption{\small Plots of the condensate as a function of the bare mass, for fixed magnetic field. Only the lowest positive branch of the condensate curve is thermodynamically stable.}
      \label{fig:fig4}
\end{figure}

To obtain our numerical results we first introduce the dimensionless radial coordinate $\tilde r =r/R$ and study numerically the function $\chi(\tilde r) =\sin\,\theta(\tilde r)$, in these coordinates the Lagrangian corresponding to (\ref{lag-H}) becomes:
\begin{equation}\label{dim-less-lag}
{\cal L}\propto \sqrt{H^2+\tilde r^2}\sqrt{1-\chi(\tilde r)^2}\,\sqrt{1-\chi(\tilde r)^2+(1+\tilde r^2)\chi'(\tilde r)^2}\ .
\end{equation}
As in the previous section we will present our results in terms of the radial coordinate $u$ (defined in equation (\ref{u_of_r})), whose dimensionless analogue is given by $\tilde  u =(\tilde r +\sqrt{1+\tilde r^2})/2$. The presence of external magnetic field does not introduce new counter terms and using the same definition for the condensate as in the previous section, we read the bare mass and the condensate of the theory from the asymptotic expansion of the field $\chi(\tilde r)$ at large~$\tilde r$:
\begin{equation}
\chi(\tilde r)=\tilde m/\tilde r+\tilde c/\tilde r^2+\dots\, ,
\end{equation}
where $\tilde m =m/R \propto m_q R_2/\sqrt{\lambda}$ and $\tilde c$ is proportional to the fundamental condensate of the theory $\langle\bar\psi\psi\rangle\propto-R^3\tilde c$. We also use that the non-linear second order differential equation for $\chi(\tilde r)$ derived from (\ref{dim-less-lag}) has predetermined boundary conditions at $\chi=1$ ($\theta=\pi/2$) and depends only on one parameter $\tilde r_0$ -- the radial distance above the origin of AdS$_5$ where the D5-brane closes. Solving perturbatively for $\chi(\tilde r)$ near $\tilde r_0$ and feeding this solution into a numerical shooting technique, we can generate all possible D5-brane embeddings. 

\begin{figure}[htbp] 
   \centering
   \includegraphics[width=3in]{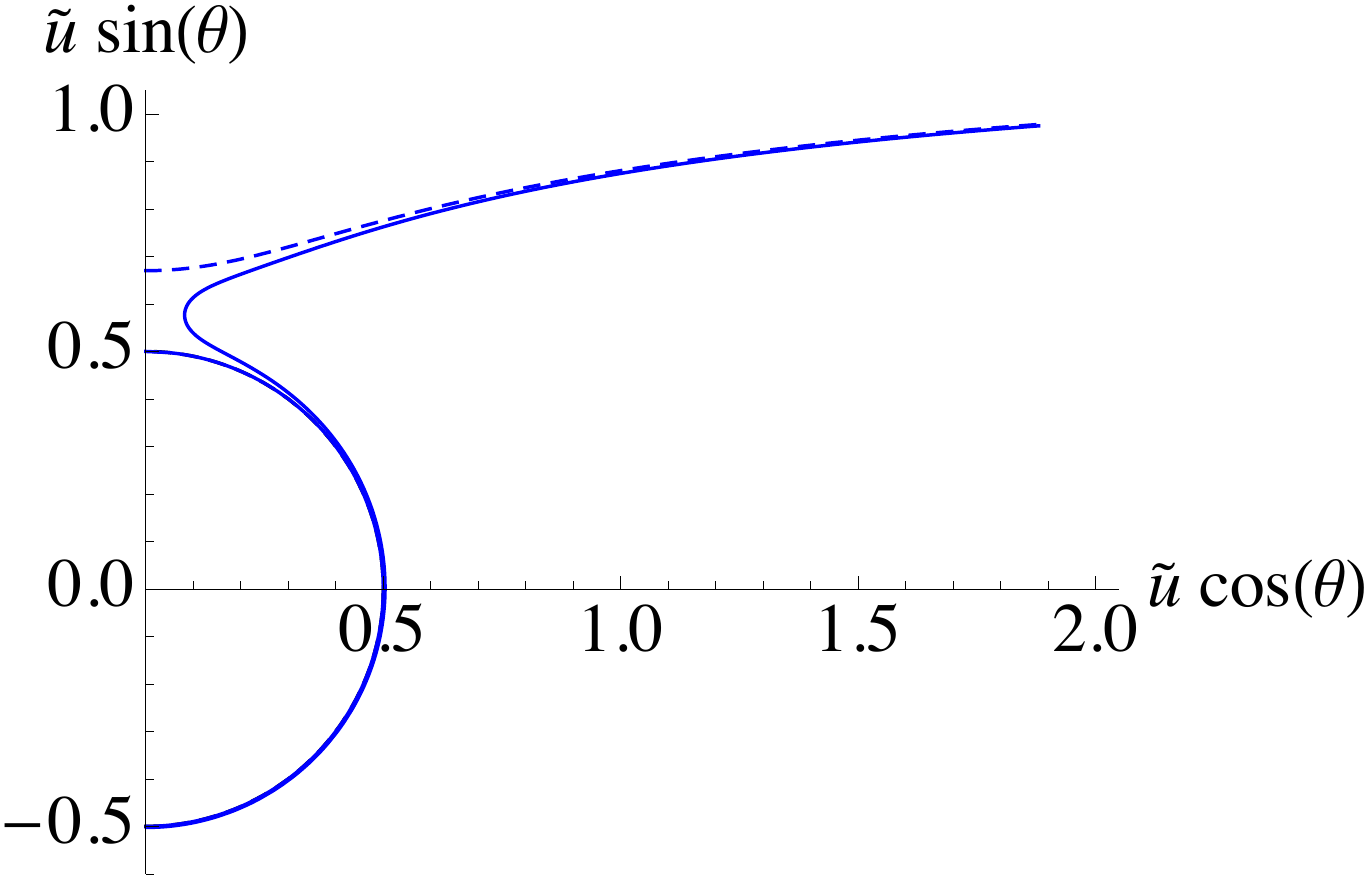} 
   \caption{The plot represents a D5-brane embedding (the solid curve) starting at $(\theta =\pi/2, r=10^{-6})$ then extending along the $S^5$ part of the geometry, folding back near $\theta=-\pi/2$ and almost reaching $\theta=\pi/2$ before entering a steady regime. In the steady regime the ``folded'' embedding approaches another D5-brane embedding (the dashed curve) originating at $(\theta =\pi/2,r=0.297)$. The two embeddings give rise to almost identical bare mass and condensate parameters.} 
      \label{fig:fig5}
\end{figure}

A plot of the condensate as a function of the bare mass for various values of the magnetic field has been presented in figure \ref{fig:fig4}. As one can see, there is a multivalued loop centred at the origin of the $(\tilde m,-\tilde c)$ plane. This loops is similar to the spiral structure, which generally appear in holographic gauge theories dual to the ``flat'' Dp/Dq intersection, when subjected to an external magnetic field \cite{Filev:2009xp}. However, unlike the scenario described in refs. \cite{Filev:2009xp},\cite{Filev:2007qu} in the present case the step of the spiral is extremely small and it seems that the spiral approaches a limiting loop (rather than shrinking to zero radius). In practice, in the plots we cannot distinguish between the spiral and the limiting loop and the curve of the condensate winds along the loop. The origin of this behaviour in the holographic set up can be understood after one parametrises the D5-brane embeddings by $\tilde r=\tilde r(\theta)$ and verifies that $\tilde r(\theta)\equiv 0$ is a fixed point of the equation of motion for $\tilde r(\theta)$. This causes embeddings, which start at infinitesimal radial distances ($\tilde r_0 \ll 1$) to bend and fold along the $S^5$ near the origin of AdS$_5$, with the function $\theta(\tilde r)$ changing rapidly form $-\pi/2$ to $\pi/2$, before they enter a steady regime of slowly varying $\theta(\tilde r)$ at some still small but larger $\tilde r_0$. In the steady regime the asymptotic properties of the D5-branes depend very weakly on changes of the initial (IR) conditions and as a result the ``folding'' solutions become extremely close to some embeddings which started at larger $\tilde r_0$ and give rise to practically the same values of the bare mass condensate at infinity. A particular case of such behaviour is plotted in figure \ref{fig:fig5}. To summarise: the condensate is a singe valued function of the bare mass at large bare masses (large initial parameter $\tilde r_0$). At very small initial parameters $r_0$ the D5-brane embeddings are affected by the fixed point $\tilde r(\theta)=0$ and start folding along the $S^5$ part of the geometry near the origin of AdS$_5$, as a result the condensate is multivalued function of the bare mass and as the initial parameter $\tilde r_0$ approaches zero the condensate curve winds clockwise along a very tight spiral approaching a limiting loop. 

Similarly to the spiral structures of refs.  \cite{Filev:2009xp},\cite{Filev:2007qu} the ``folding'' solutions have higher free energy (they are also unstable) and only the first (lower) branch of the spiral remains. Furthermore, since negative bare masses can be mapped to positive bare masses with flipped condensate, one can also show that the (lower) positive branch of the condensate curve has the lowest free energy this is why we focus on its properties. 

Finally we point out that at zero bare mass the condensate has a finite negative value (for the stable branch) suggesting that a spontaneous breaking of the global symmetry corresponding to rotations along the transverse two sphere is taking place. Therefore our system realises the effect of magnetic catalysis of spontaneous symmetry breaking \cite{mag-cat}.

\begin{figure}[htbp] 
   \centering
   \includegraphics[width=2.7in]{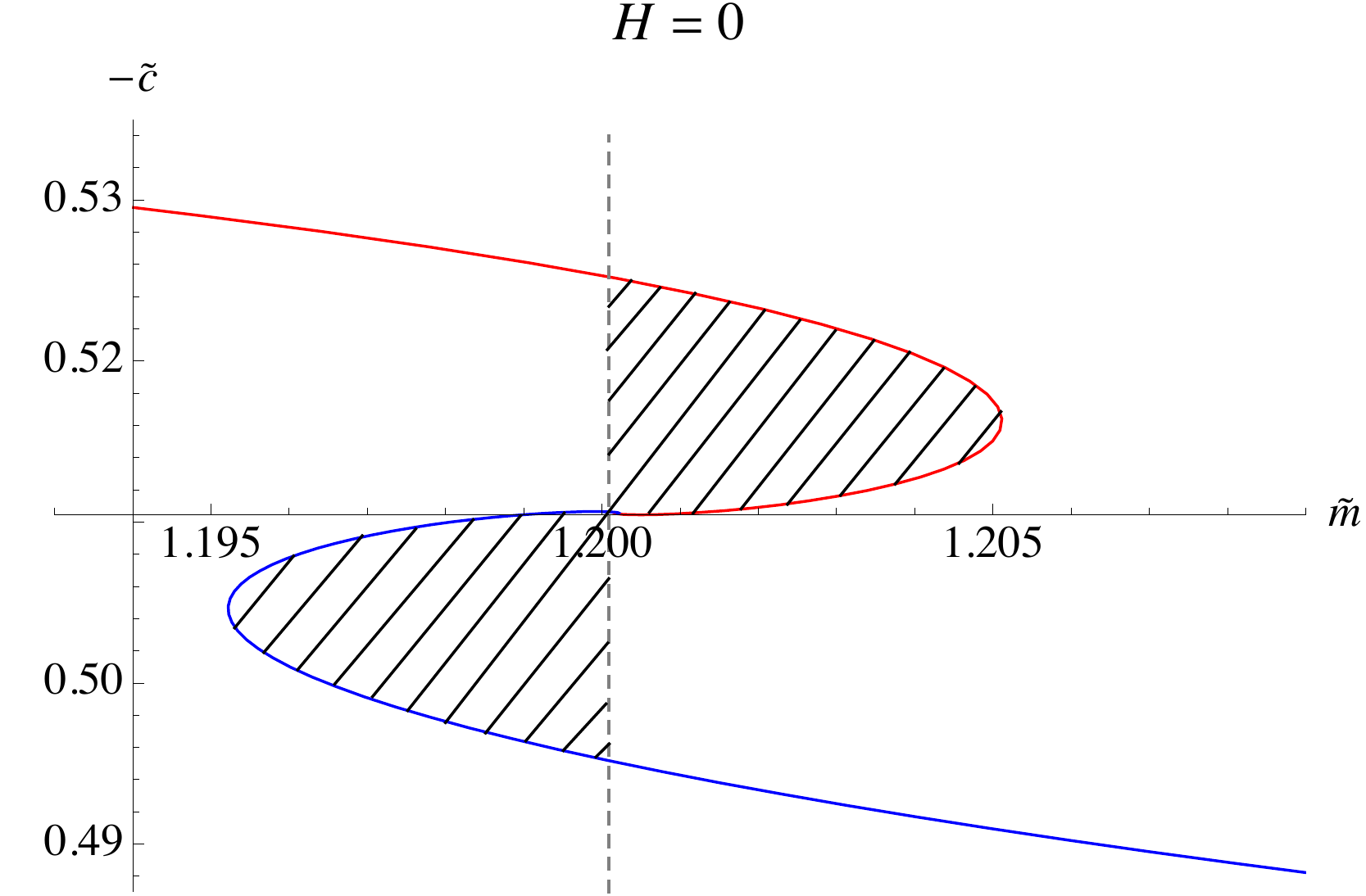} 
    \includegraphics[width=2.7in]{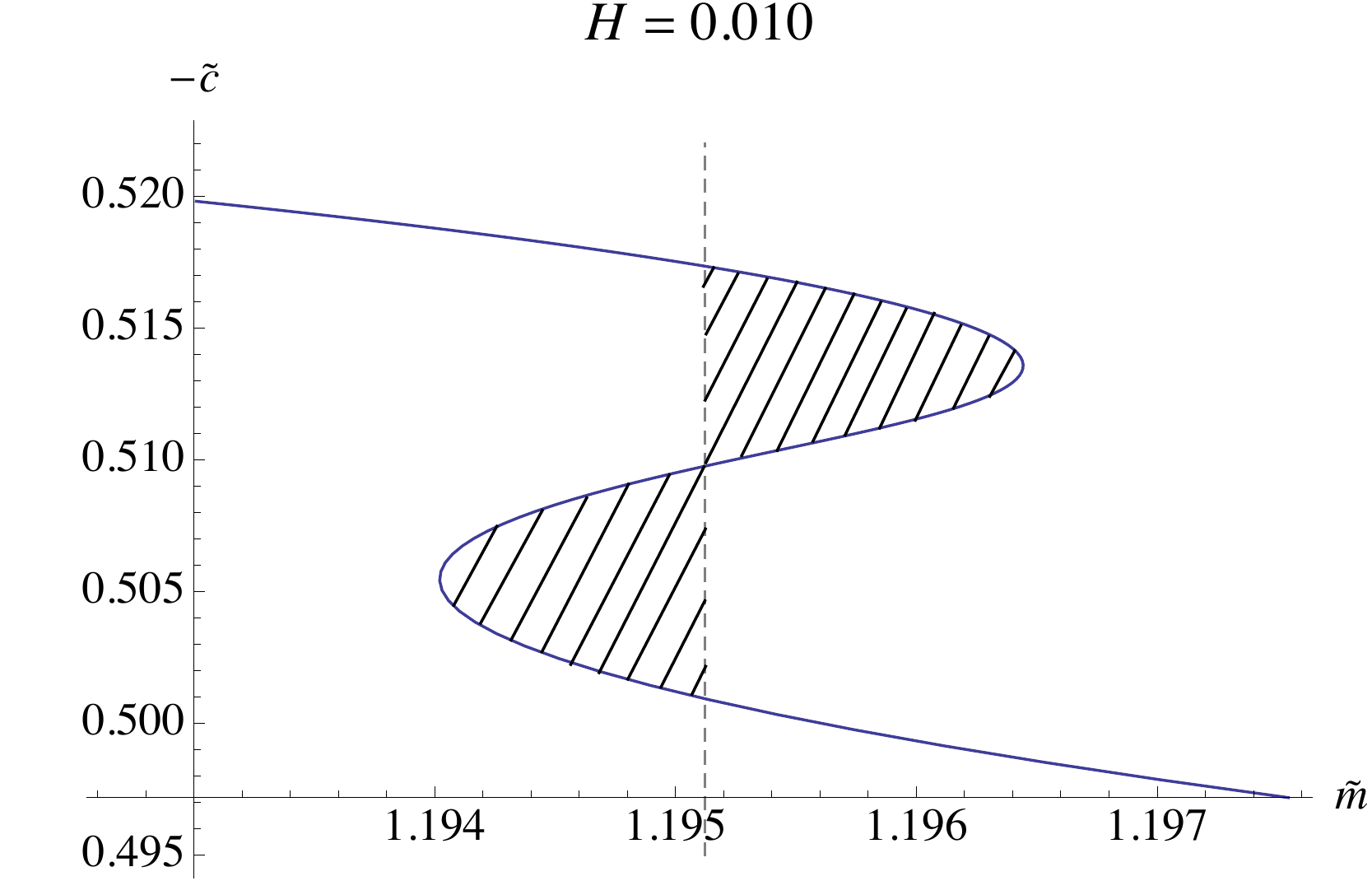} 
    \vspace{0.1cm}
    \includegraphics[width=2.7in]{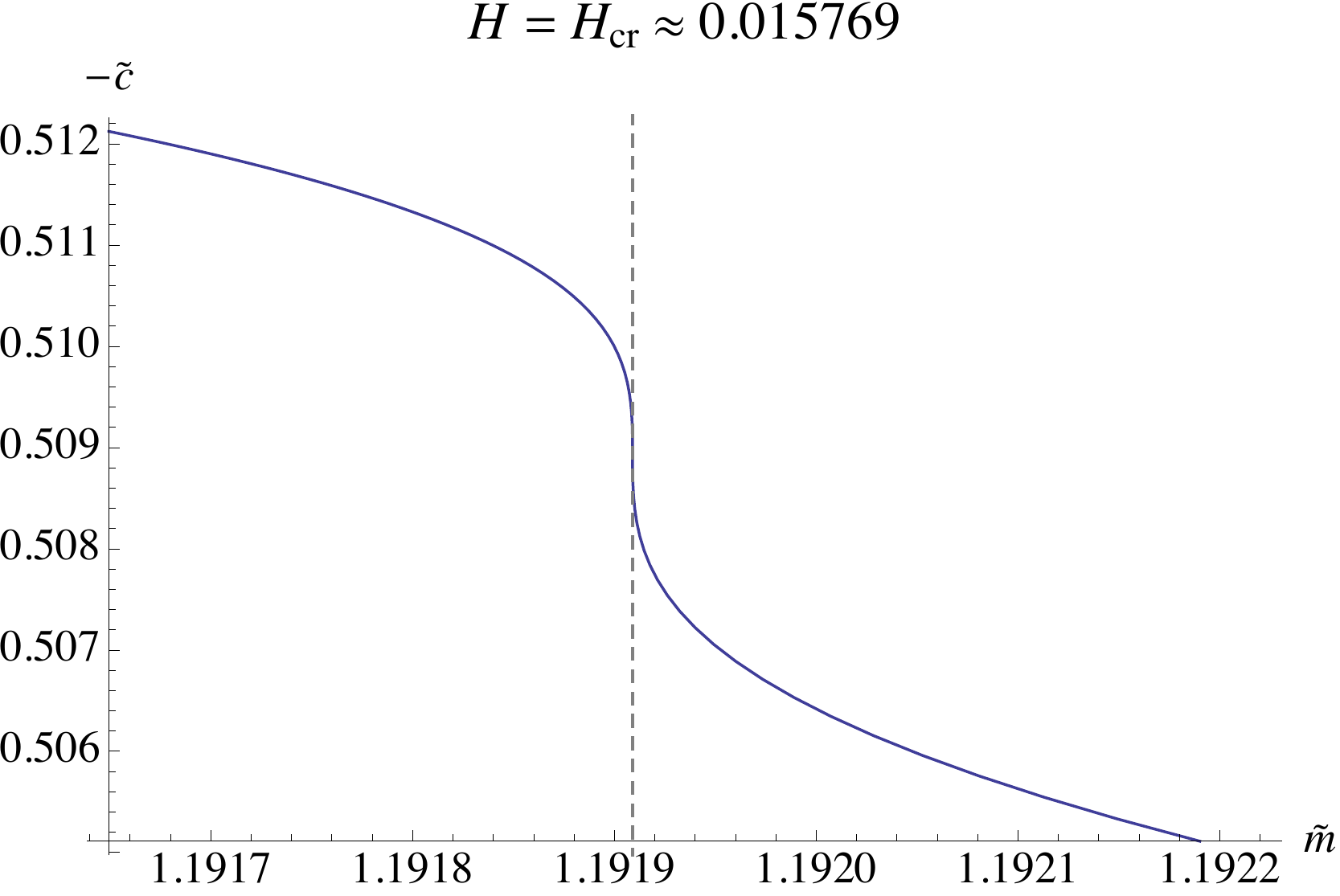}
    \includegraphics[width=2.7in]{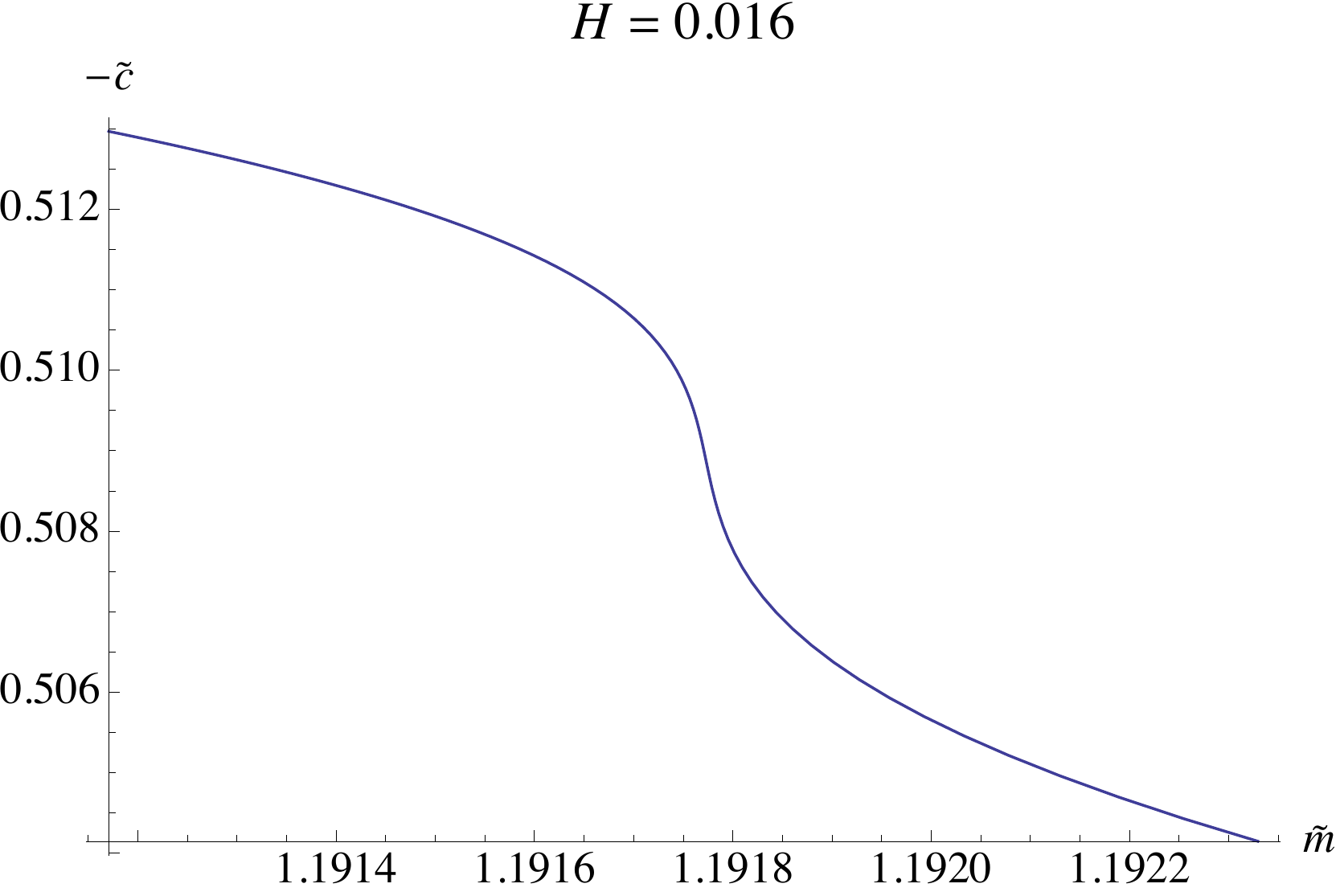}  
   \caption{\small Plots of the condensate versus bare mass parameters zoomed in near the phase transition. One can see that for $H < H_{cr}$ there is a phase transition within the confined phase mimicking the phase transition at $H=0$, at $H=H_{cr}$ the phase transition ends on a critical point of a second order phase transition. For $H >H_{cr}$ the phase transition is replaced by a smooth crossover.} 
      \label{fig:fig6}
\end{figure}
\subsection{Phase Diagram}
The most interesting property of the condensate curve is the existence of a first order phase transition pattern for sufficiently small magnetic fields, which ends at a critical point of a second order quantum phase transition for $H=H_{cr}\approx 0.015769$. In figure \ref{fig:fig6} we have presented plots of the condensate zoomed in near the region of the phase transition. One can see that for for $H < H_{cr}$ there is a first order phase transition within the confined phase, which mimics the confinement/deconfinement phase transition at vanishing magnetic field. The critical parameter $\tilde m_{cr}$, at which the phase transition takes place, can be found using Maxwell's equal are law, or alternatively by computing the density of the regularised wick rotated on-shell action, which is identified with the free energy  density of the system.  The two approaches give the same result, because the condensate is a derivative of the free energy density with respect to the bare mass (at fixed radius of the two sphere and fixed magnetic field). One can also see that at $H=H_{cr}$ the phase transition ends on a critical point of a second order phase transition and for $H>H_{cr}$ the phase transition is replaced by a smooth cross over. 

\begin{figure}[t] 
   \centering
   \includegraphics[width=4.5in]{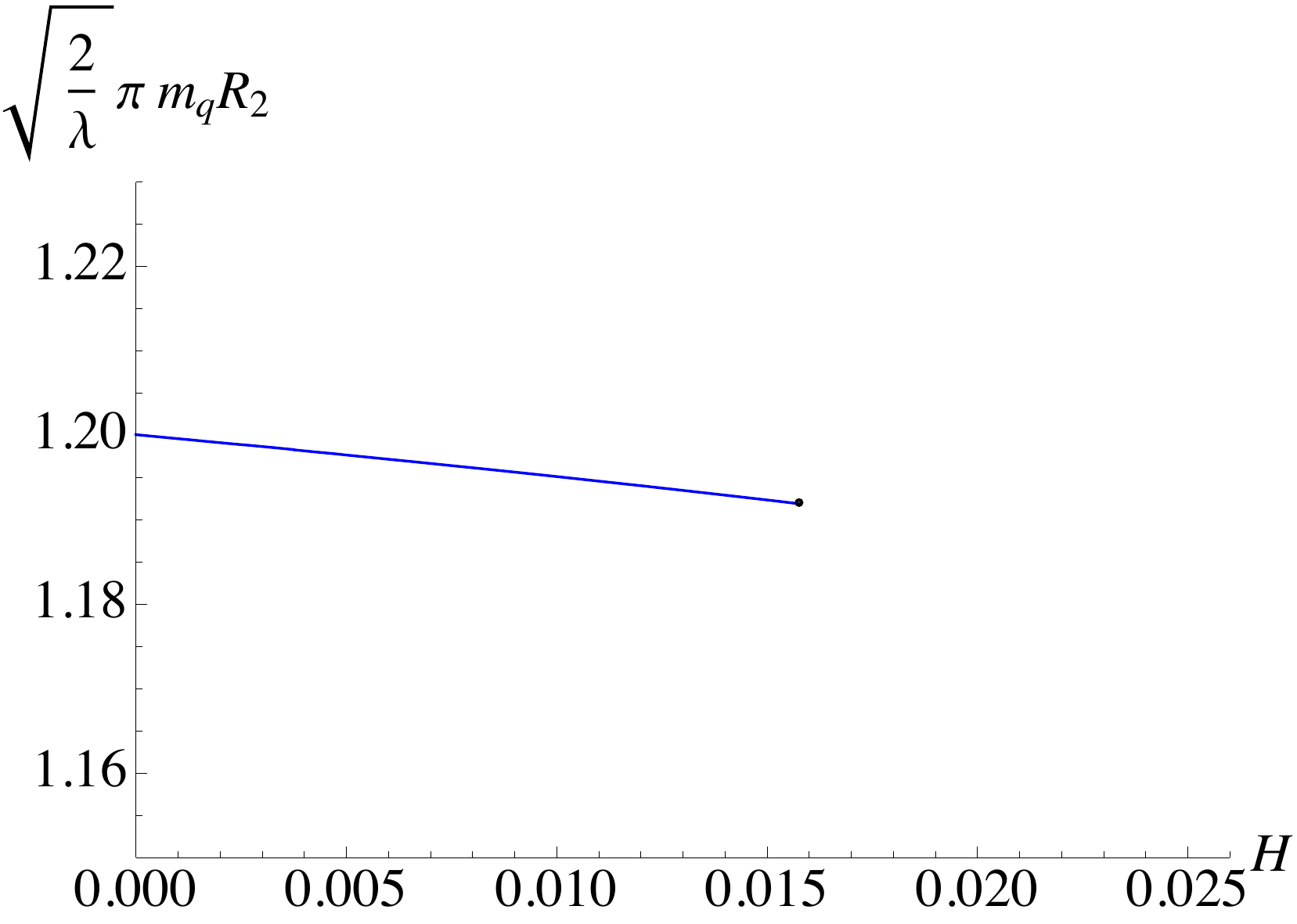} 
   \caption{\small A plot of the phase diagram of the theory. The Blue curve represents a critical line of a first order phase transition. The critical line ends at a critical point of a second order phase transition (the solid dot).} 
      \label{fig:fig7}
\end{figure}

The observations from figure \ref{fig:fig6} can be summarised in the phase diagram presented in figure \ref{fig:fig7}, where we have presented the bare mass parameter in physical notations $\tilde m =\sqrt{\frac{2\pi^2}{\lambda}}\,m_q\,R_2$. The blue curve in the figure represents a critical line of a first order phase transition. The critical line ends on a critical point of a second order phase transition (the solid dot). In the next subsection we focus on the properties of the theory at this critical point. 
\subsection{Magnetisation}
\begin{figure}[t] 
   \centering
   \includegraphics[width=2.72in]{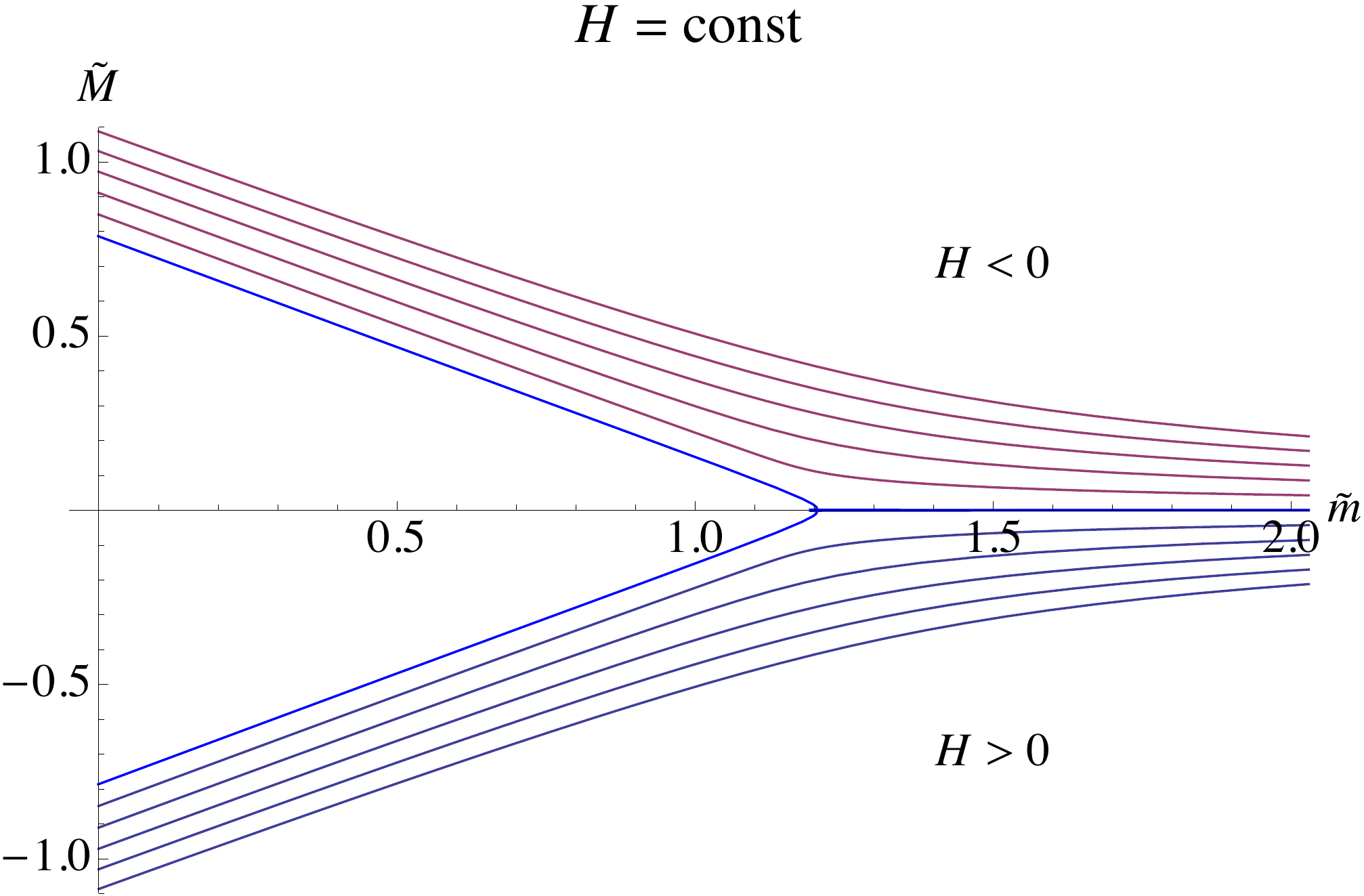} 
   \includegraphics[width=2.72in]{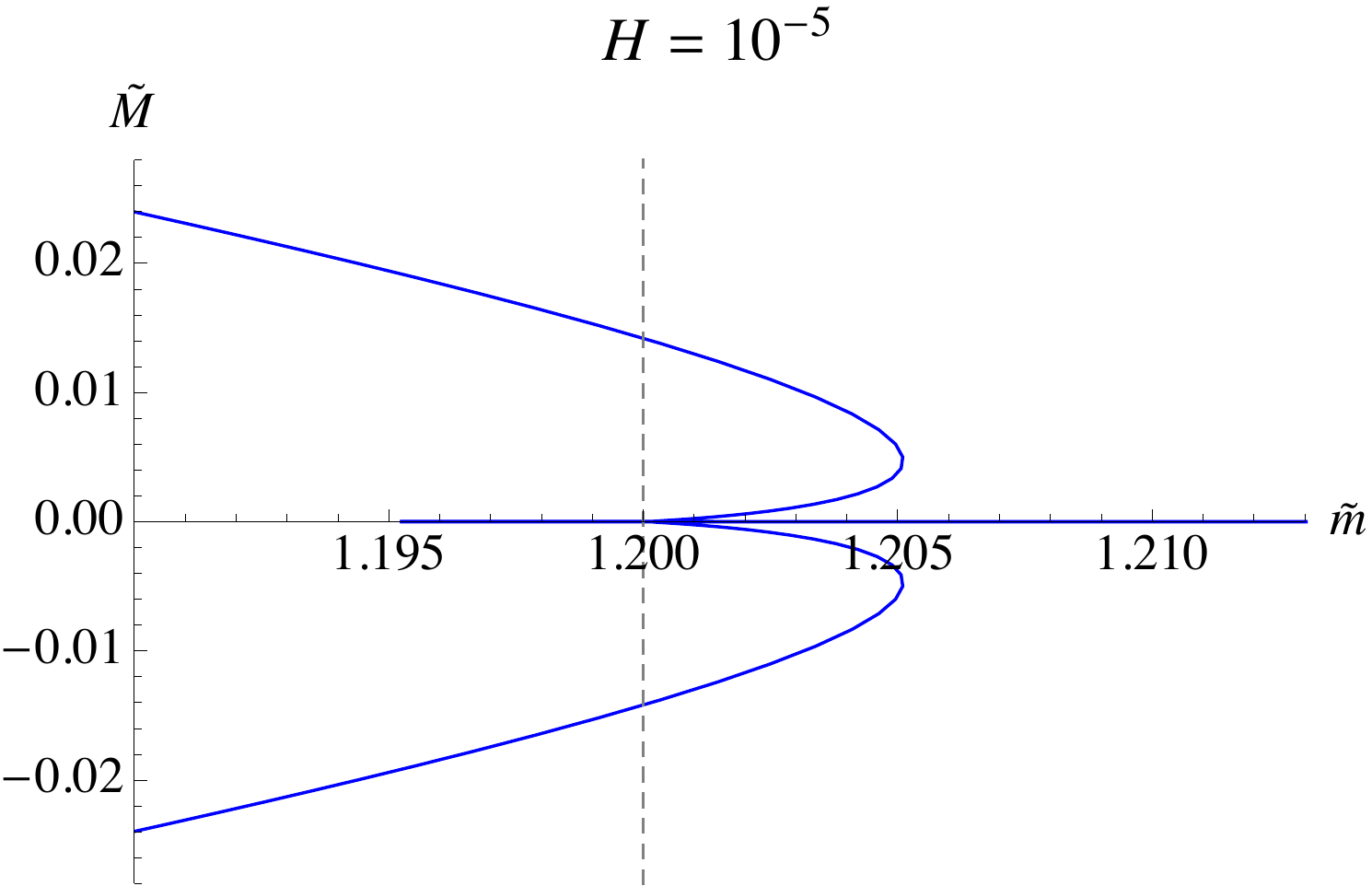} 
   \caption{\small \emph{To the left:} Plots of the magnetisation as a function of the bare mass parameter $\tilde m$ for fixed values of the magnetic field ranging from $|H|=10^{-3}$ to $|H|=0.30$. The innermost (blue) curve corresponding to $|H|=10^{-3}$ is representative for the limits $H\to 0^+$ and $H\to 0^-$ and one can see that for $\tilde m \leq m_*\approx 1.20$ the magnetisation remains finite suggesting that theory has a persistent diamagnetic response in this regime. \emph{To the right:} A zoom of the critical region. One can see that at $\tilde m =\tilde m_*$ there is a first order phase transition from a ``persistent'' diamagnetism to a state with vanishing magnetisation.}
      \label{fig:fig8}
\end{figure}

In this subsection we study the magnetisation of the system. Identifying the free energy (spacial) density with the ($2+1$ dimensional) density of the regularised euclidean ``on-shell'' action, for the free energy density we obtain:
\begin{eqnarray}\nonumber
\frac{F}{\cal N'}&=&\int_{\tilde r_0}^{\infty}d\tilde r\left(\sqrt{\tilde r^4+H^2}\,\sqrt{1-\chi(\tilde r)^2}\,\sqrt{1-\chi(\tilde r)^2+(1+\tilde r^2)\chi'(\tilde r)^2}-\tilde r^2+\frac{\tilde m^2}{2}\right)+\\
&-&\frac{\tilde r_0^3}{3}+\frac{\tilde m^2}{2}\tilde r_0+\tilde c\,\tilde m\ ,
\label{Free-energy}
\end{eqnarray}
where ${\cal N'}={\mu_5\,2\pi R^3}/{g_s}$, we used that the volume of the field theory $S^2$ is $2\pi R^2$ and we subtracted the counter terms prescribed in refs. \cite{Karch:2009ph}, \cite{Karch:2005ms}. Using equation (\ref{Free-energy}) and that $M=-{\partial F}/{\partial H_{\rm phys}}|_{m_q}$, one easily arrives at the following expression for the magnetisation:
\begin{equation}\label{tilde-M}
\frac{M}{2\pi\alpha'\,{\cal N'}}=\tilde M=-H\,\int_{\tilde r_0}^{\infty}d\tilde r\, \frac{\sqrt{1-\chi(\tilde r)^2}\,\sqrt{1-\chi(\tilde r)^2+(1+\tilde r^2)\chi'(\tilde r)^2}}{\sqrt{\tilde r^4+H^2}}\ ,
\end{equation}
where we have introduced the dimensionless quantity $\tilde M = {M}/{(2\pi\alpha'\,{\cal N'})}$. In figure \ref{fig:fig8} we have presented a plot of the parameter $\tilde M$ at fixed magnetic field as a function of the bare mass parameter $\tilde m$. The plots represent decreasing values of the magnetic field from $|H|=0.30$ to $|H|=10^{-3}$. Note that the magnetisation is negative (for positive $H$) and hence the system is diamagnetic. The innermost (blue) curve corresponds to $|H|=10^{-3}$ and is in practise indistinguishable in the plot from the curve corresponding to the limit $H\to 0^+$. The most interesting property of this curve is that for $\tilde m \leq \tilde m_* \approx 1.20$ the magnetisation has a non-zero value\footnote{In the holographic set up the source of the finite magnetisation in the limit of vanishing magnetic field is the D3-brane throat whose tension is proportional to $H$ (see (\ref{DBI-D3})).} in the limits $H\to 0^+$ and $H\to 0^-$ (the lower and upper branch of the blue curve). Therefore, in this regime the theory has a persistent diamagnetic response, which is independent of the magnetic field. This effect is very similar to the effect of persistent diamagnetic current in mesoscopic systems, such as nano tubes and quantum dots \cite{Zelyak}. Although such systems contain large namer of microscopic elements (electrons, molecules, etc.) so that macroscopic quantities are well defined, they are still relatively small and experience strong finite size quantum effects, which are the main cause of their peculiar properties. Recall that the bare mass parameter $\tilde m$ is proportional to the radius of the field theory two sphere $R_2$ and hence to the size of the system. Therefore, the persistent diamagnetic response is realised when the systems is sufficiently small ($\tilde m < m_*$) and finite size quantum effects (such as the existence of Casimir energy) are important, this makes the analogy to mesoscopic systems even stronger. 
As can be seen from the second plot in figure \ref{fig:fig8} the phase transition from ordinary to ``persistent'' diamagnetism is of a first order, where the critical parameter $\tilde m_*\approx 1.20$ has the same value as for the confinement/deconfinement phase transition at strictly vanishing magnetic field reviewed in section \ref{sec2}.
Another interesting property of the magnetisation is its behaviour near the quantum critical point. Plots of the magnetisation $\tilde M$ and the corresponding magnetic susceptibility $\tilde \chi_m =\partial\tilde M/\partial H$ as functions of the magnetic field and for $\tilde m =\tilde m_{cr}$ are presented in figure {\ref{fig:fig9}. From the second plot one can see that the magnetic susceptibility diverges at the phase transition, as expected for a second order phase transition. The question for the critical exponents that govern this behaviour will be addressed in the next subsection where we focus on the critical regime of the theory.
\begin{figure}[t] 
   \centering
\hspace{-.7em}   \includegraphics[width=2.7in]{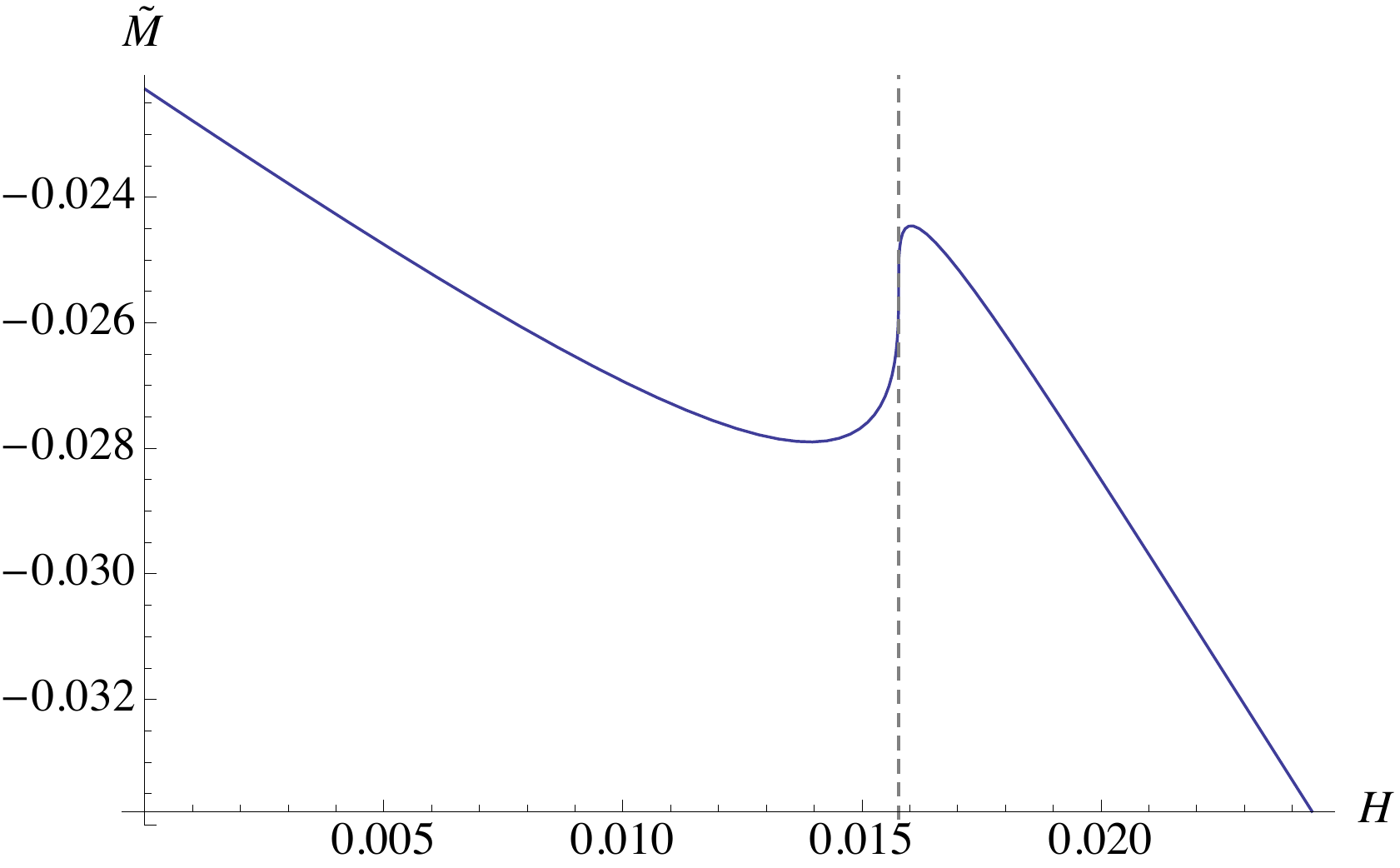} 
 \hspace{1.4em}    \includegraphics[width=2.7in]{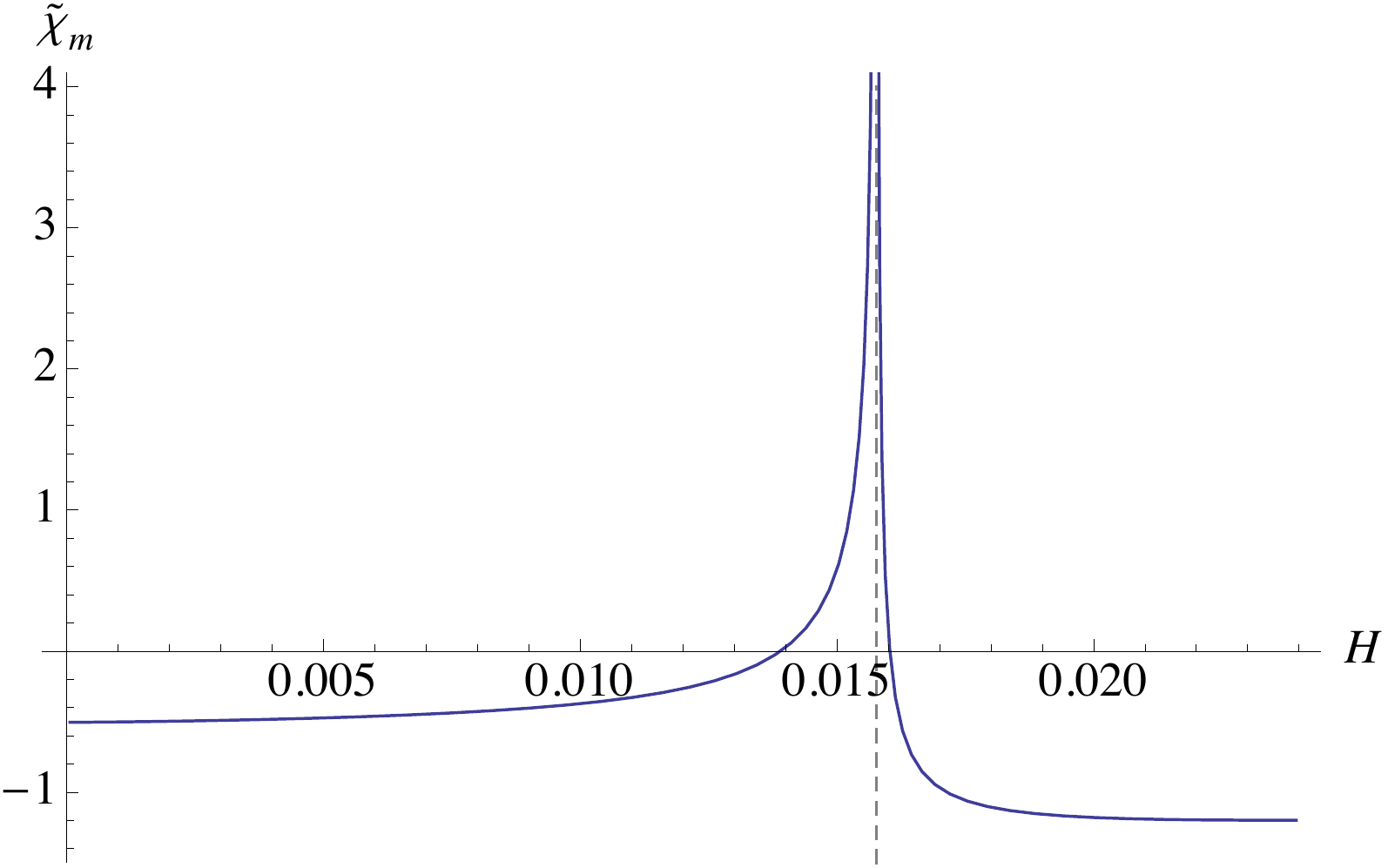} 
   \caption{\small Plots of the magnetisation $\tilde M$ and the magnetic susceptibility $\tilde \chi_m =\partial\tilde M/\partial H$ as functions of the magnetic field at fixed bare mass parameter $\tilde m=\tilde m_{cr}$. As expected the susceptibility diverges at the phase transition.}
      \label{fig:fig9}
\end{figure}

\subsection{Critical behaviour}\label{crit-exp}

In this subsection we explore the properties of the theory at the critical point, in particular we calculate the critical exponents of various observables. For all observables that we study we find with a very high numerical accuracy that the critical exponent is one third, and for the corresponding diverging susceptibilities we find critical exponent $-2/3$. Generally, it is expected that at a second order phase transition the second derivatives of the free energy will diverge. In our considerations we will keep the radius of the two sphere fixed. The variation of the free energy density is then given by:
\begin{equation}
dF = -M \,dH_{\rm phys} +\langle\bar\psi\psi\rangle\,  dm_q\ ,
\end{equation}
where $M$ is the magnetisation of the theory and $H_{\rm phys} =H/(2\pi\alpha')$ is the physical magnetic~field. 
\begin{figure}[t] 
   \centering
   \includegraphics[width=3.7in]{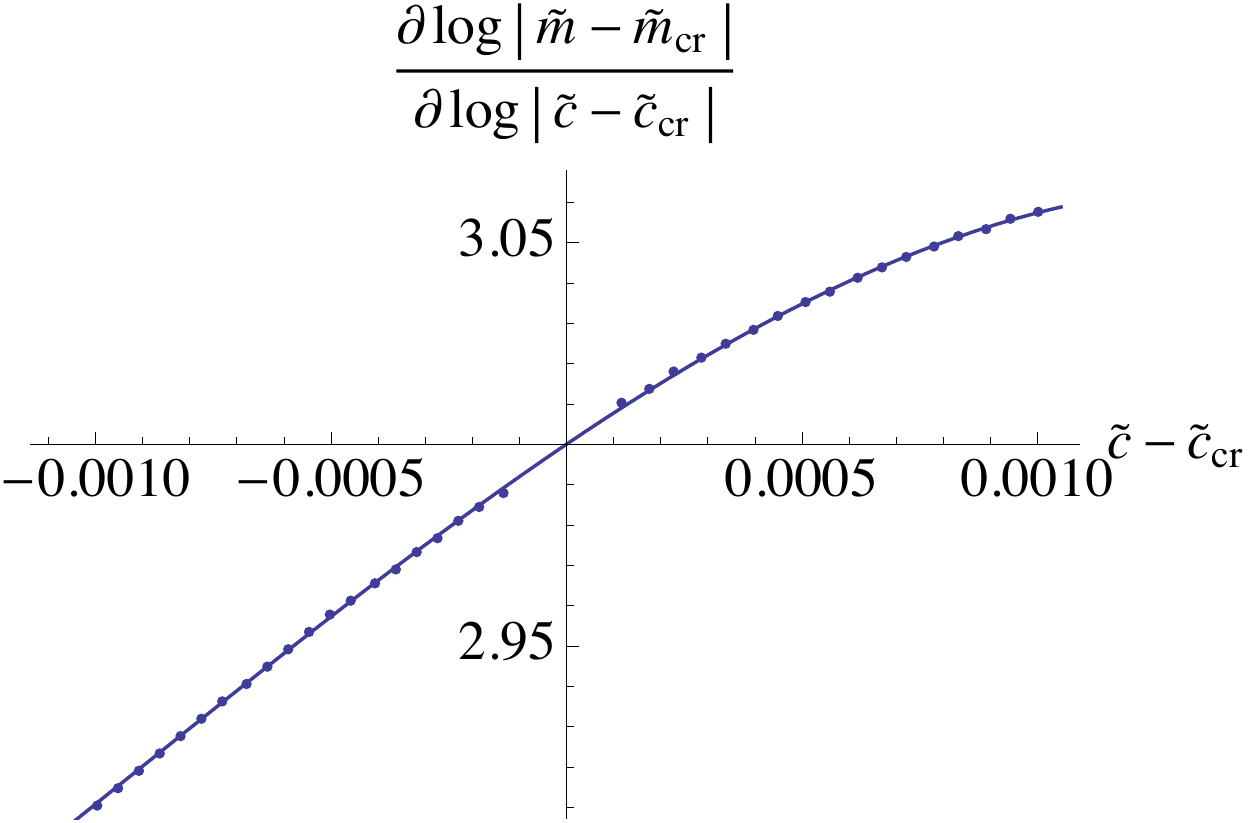} 
   \caption{\small A plot of the logarithmic derivative  $\partial \log|\tilde m-\tilde m_{cr}|/\partial \log|\tilde c -\tilde c_{cr}|$ as a function of $\tilde c-\tilde c_{cr}$. The solid curve represents a non-linear fit to the numerical data. } 
      \label{fig:fig10}
\end{figure}
\begin{enumerate}[label=(\alph*)]
 \item 
 We begin by studying the second derivative of the free energy with respect to the bare mass  $m_q$ at fixed magnetic field  $\partial^2 F/\partial m_q^2=\partial \langle\bar\psi\psi\rangle/\partial m_q$. At a fixed radius of the two sphere we have: $\langle\bar\psi\psi\rangle\propto -\tilde c$ and $m_q\propto \tilde m$, and it is sufficient to study the behaviour of the derivative $-\partial \tilde c/\partial\tilde m$ at a fixed magnetic field ($H=H_{cr}$). We will show that as the bare mass parameter approaches the critical value $\tilde m_{cr}\approx 1.9191$ the first derivative of the condensate diverges as $|\tilde m-\tilde m_{cr}|^{\gamma-1}$, where $\gamma$ is the critical exponent characterising the behaviour of the condensate near the phase transition, that is near the phase transition we have:
\begin{equation}
 |\tilde c-\tilde c_{cr} | =|\tilde m-\tilde m_{cr}|^{\gamma}\ .
\end{equation}
 Note that in general the crucial exponents on both sides of the transition need not agree, however this is often the case in critical phenomena and our system is not an exception. To determine numerically the value of the critical exponent $\gamma$ we study the logarithmic derivative $\partial \log|\tilde c-\tilde c_{cr}|/\partial \log|\tilde m -\tilde m_{cr}|$ whose value in the limit $\tilde m\to\tilde m_{cr}$ coincides with $\gamma$. Refining our numerical techniques and focusing on the region near the phase transition, we obtain the data presented in figure \ref{fig:fig10}.

Note that the points in a very close proximity to the phase transition have been removed to avoid the numerical evaluation of a term behaving like $0/0$,  which would inevitably generate a huge numerical error. Note also that we have evaluated $1/\gamma$. A non-linear fit (the solid curve) to the data, provided a value for the critical exponent of $1/\gamma =3.0000\pm 0.0002$ with 95\% confidence. We consider this result as a very strong indication that the critical exponent is $\gamma=1/3$ and the second derivative of the condensate diverges at the phase transition as:
 \begin{equation} 
\frac{\partial^2F}{\partial m^2}\propto|\tilde m-\tilde m_{cr}|^{-2/3}.   
\end{equation}

\item Our next focus is the magnetic susceptibility $\chi_m$ related to the second derivative of the free energy with respect to the magnetic field: $\chi_m=\partial M/\partial H_{\rm phys}=-\partial^2 F/\partial H_{\rm phys}^2$. 
\begin{figure}[t] 
   \centering
\hspace{-.7em}   \includegraphics[width=2.7in]{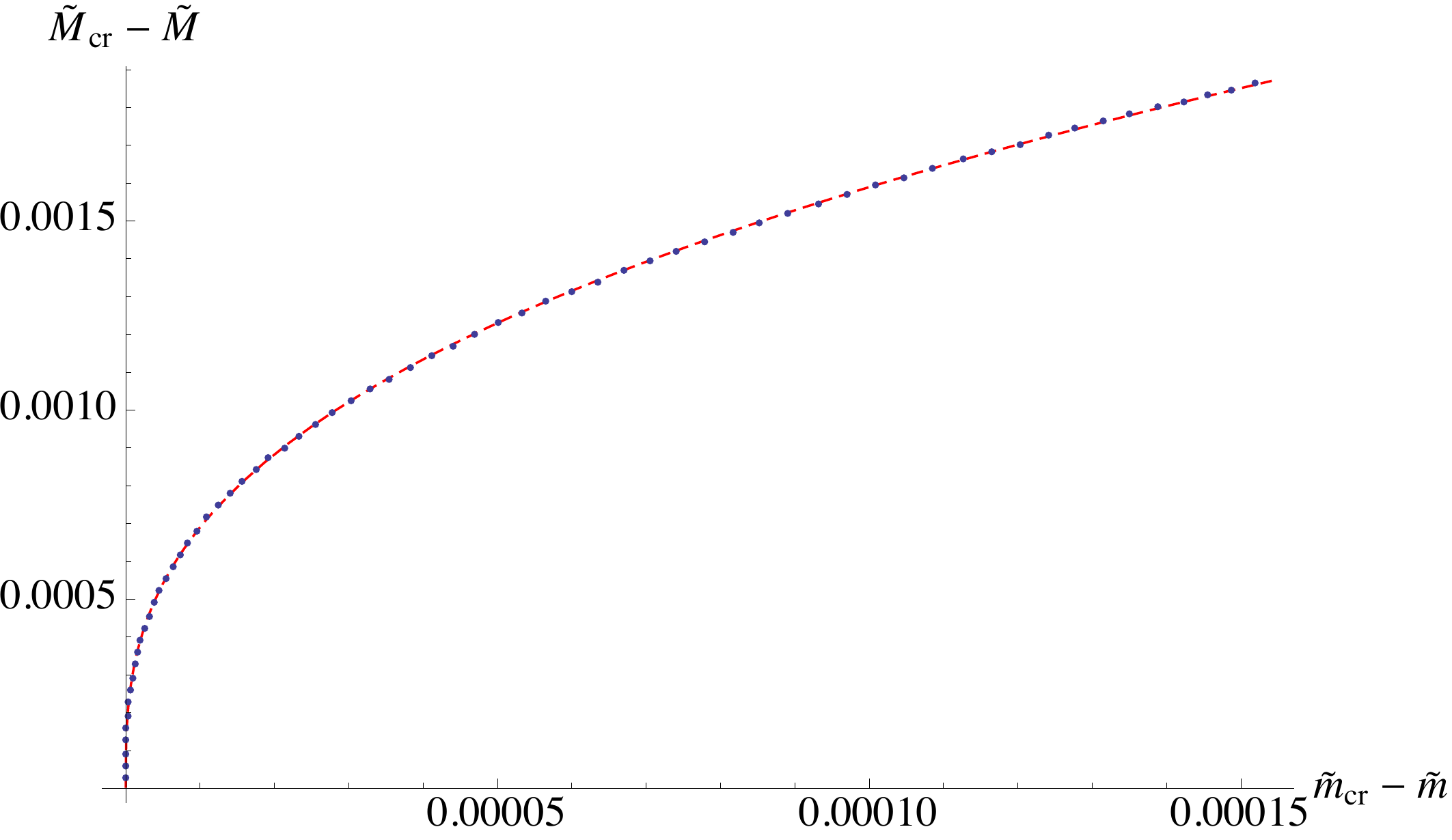} 
\hspace{1.4em}   \includegraphics[width=2.7in]{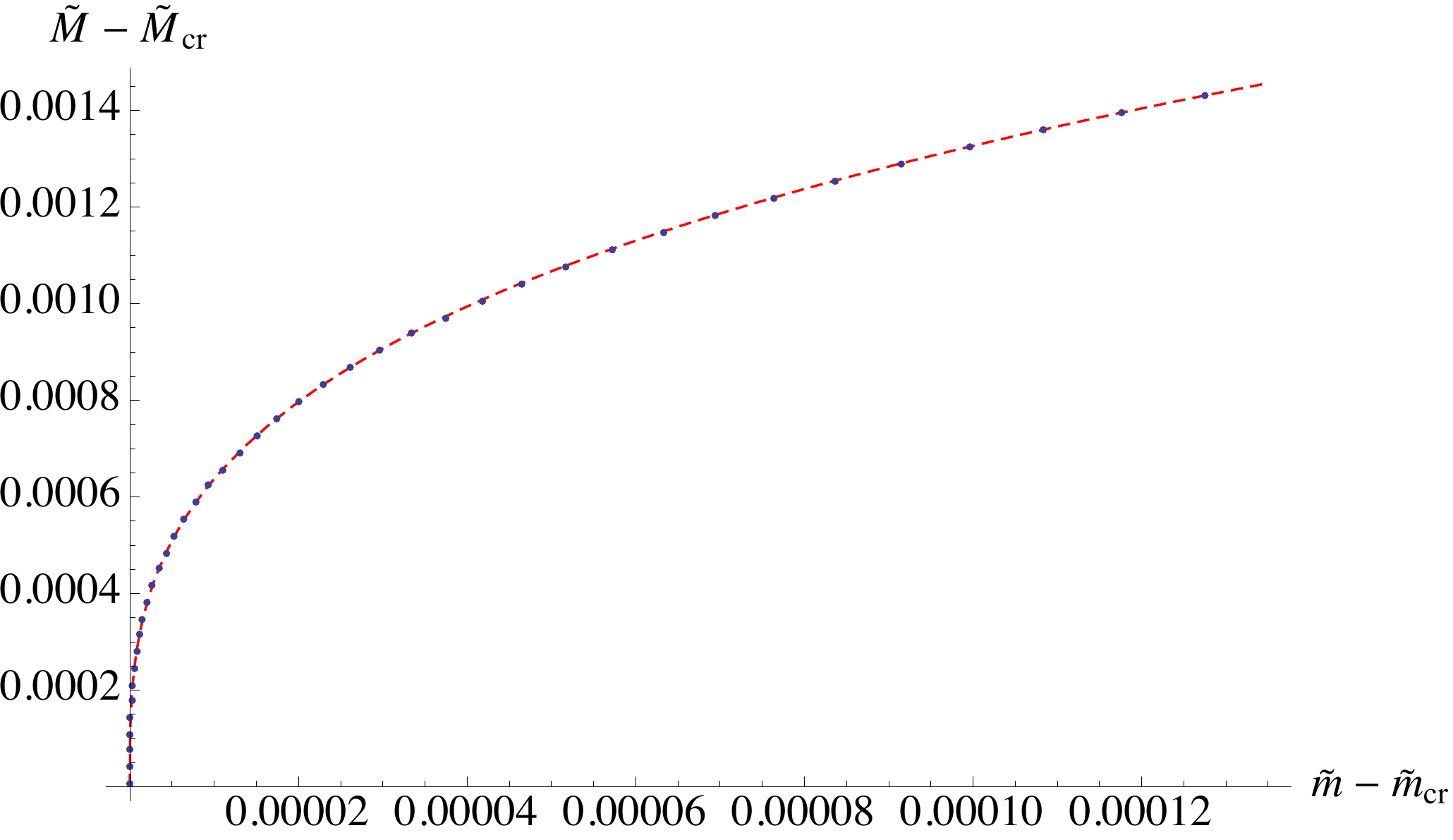} 
   \caption{\small Plots of the quantities $\pm(\tilde M-\tilde M_{cr})$ versus $\pm(\tilde m-\tilde m_{cr})$ at $H=H_{cr}$, the dashed curves represent ${\rm const}_1\, x^{1/3} +{\rm const}_2\, x^{2/3}$ fits.}
      \label{fig:fig12}
\end{figure}
For numerical studies it is convenient to introduce the dimensionless quantity $\tilde\chi_m =\partial\tilde M/\partial H$, where $\tilde M$ is defined in equation (\ref{tilde-M}). Plots of the parameters $\tilde M$ and $\tilde \chi_m$ have already been presented in figure \ref{fig:fig9}. In this subsection we are interested in the behaviour of the parameters across the phase transition. This is why we generate a plot of the magnetisation at $\tilde m=\tilde m_{cr}$ for a small range of the magnetic field $H$ around the critical value: $H_{cr} -\Delta H \leq H\leq H_{cr}+\Delta H$. In figure \ref{fig:fig11} we have presented plots of the quantities $\pm(\tilde M-\tilde M_{cr})$ versus $\pm(H-H_{cr})$, the dashed curves represent ${\rm const}_1\, x^{1/3} +{\rm const}_2\, x^{2/3}$ fits. One can see the excellent agreement on both sides of the phase transition, suggesting that the magnetisation approaches its critical value as $\tilde M-\tilde M_{cr}\propto (H-H_{cr})^{1/3}$ and hence for the critical behaviour of the susceptibility we obtain:
\begin{equation}
\chi_m\propto |H-H_{cr}|^{-2/3}\ .
\end{equation}
We see that the magnetic susceptibility has the same critical exponent as the second derivative of the free energy with respect to the bare mass $\partial^2 F/\partial m^2$.

\item Finally, we are interested in the properties of the mixed derivative $\partial^2 F/\partial m\,\partial H = -\partial M/\partial m$ across the phase transition. To study its behaviour we generate a plot of the magnetisation parameter $\tilde M$ at $H=H_{cr}$ and for a small range of the bare mass parameter $\tilde m$ around the critical one: $\tilde m_{cr}-\Delta\tilde m \leq\tilde m \leq \tilde m_{cr}+\Delta \tilde m$.  As can be seen from the plots in figure \ref{fig:fig12}, there is an excellent agreement with the ${\rm const}_1\, x^{1/3} +{\rm const}_2\, x^{2/3}$ fits, suggesting that the second derivative $\partial^2 F/\partial m\,\partial H$ diverges as:
\begin{equation}
\frac{\partial^2 F}{\partial m\,\partial H}\propto |\tilde m-\tilde m_{cr}|^{-2/3}
\end{equation}
across the phase transition. We will return to the critical behaviour of the theory in the next section when we study the behaviour of the meson spectrum of the theory near the phase transition and show that at the phase transition the theory has a massless mode, which is also approached with a critical exponent $1/3$.
\end{enumerate}

\section{Meson spectrum}\label{sec4}

In this section we analyse the spectrum of the quadratic fluctuations of the system corresponding to the spectrum of the mesonic bound states in the dual field theory. To obtain the equations of motion for the fluctuations we expand near the classical profile of the D5-brane:
\begin{figure}[t]
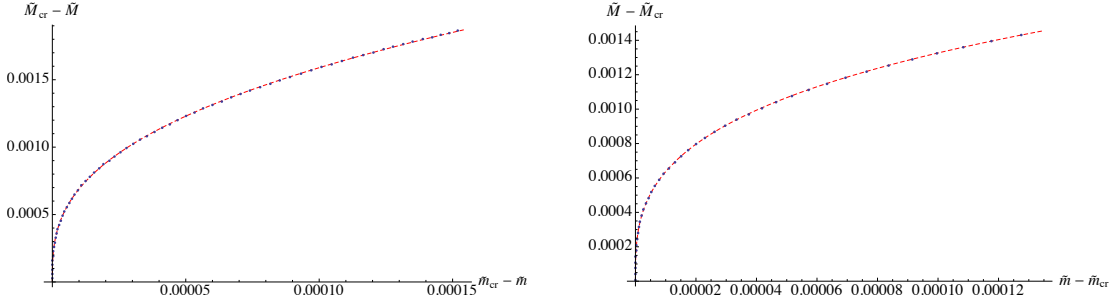
 
   \centering
\hspace{-0.7em}   \includegraphics[width=2.7in]{MagZoom1_vs_m.pdf} 
 \hspace{1.4em}  \includegraphics[width=2.7in]{MagZoom2_vs_m.pdf} 
   \caption{\small Plots of the quantities $\pm(\tilde M-\tilde M_{cr})$ versus $\pm(\tilde m-\tilde m_{cr})$ at $H=H_{cr}$, the dashed curves represent ${\rm const}_1\, x^{1/3} +{\rm const}_2\, x^{2/3}$ fits.}
      \label{fig:fig12}
\end{figure}
\begin{eqnarray}\nonumber
\beta &=&\pi/2+(2\pi\alpha')\delta \beta;~~ \chi =\chi_0(r)+(2\pi\alpha')\delta \chi;~~\hat\alpha=\pi/2+(2\pi\alpha')\delta \hat\alpha;\\
\hat\gamma&=&0+(2\pi\alpha')\delta \hat\gamma\;~~~A_{\mu}=0+\delta A_{\mu}\ ,
\end{eqnarray}
where $\hat\alpha, \hat\gamma$ parameterise the $S^2$ transverse to the D5-brane and the perturbations are allowed to depend on all word volume coordinates. The next step is to expand to second order in $\alpha'$ the total action of the D5-brane:
\begin{equation}
S_{\rm D5} =-\frac{\mu_5}{g_s}\int d^6\xi\,e^{-\Phi}\sqrt{-{\rm det}(P[G+B]+2\pi\alpha'F)}+\mu_5\int \sum_kP[C_k\wedge e^{2\pi\alpha'\cal F}]\ ,
\end{equation}
where $2\pi\alpha'{\cal F}=B+2\pi\alpha'F$. Using the equations of motions for the classical embedding of the D5-brane, one can show that the first order term in the expansion of the action vanishes, as it should since we are expanding near a local extremum. Defining the matrix $E=P[G+B]$ and expanding $E =E_0+(2\pi\alpha')E_1+(2\pi\alpha')^2E_2$, for the second order expansion of the DBI action quite generally we obtain: 
\begin{equation}\label{S2DBI}
\frac{S_{\rm DBI}^{(2)}}{(2\pi\alpha')^2}=-\frac{\mu_5}{2\,g_s}\int d^6\xi\sqrt{-|E_0|}\,\left[{\rm Tr}\left(E_0^{-1}E_2-\frac{1}{2}(E_0^{-1}E_1E_0^{-1}E_1) \right)+\frac{1}{4}\left({\rm Tr}\,E_0^{-1}E_1\right)^2\right]\ .
\end{equation}
On the other hand, since the supergravity background that we consider has a non-zero RR potential $C_{(4)}$, for the second order term of the Wess--Zumino action we obtain:
\begin{equation}\label{S2WZ}
\frac{S_{\rm WZ}^{(2)}}{(2\pi\alpha')^2}=\frac{\mu_5}{(2\pi\alpha')^2}\int P[C_{(4)}]\wedge B=-\frac{4\mu_5R^6}{g_s}\int d^6\xi\sin\alpha\sin\tilde\alpha\,\chi_0(r)^2(1-\chi_0(r)^2)\delta\hat\gamma\,\partial_t\delta\hat\alpha\ ,
\end{equation}
where we have used a gauge in which the magnetic component of the background RR four potential is given by:
\begin{equation}
\tilde C_{(4)}=\frac{1}{g_s}4R^4\hat\gamma\,\sin^2 2\theta\,\sin\tilde\alpha\,\sin\hat\alpha \,d\theta\wedge d\tilde\alpha\wedge d\tilde\gamma\wedge d\hat\alpha\ ,
\end{equation}
and that $\chi_0(r)=\cos\theta(r)$. We see that the Wess--Zumino part of the action couples the fluctuations along $\hat\alpha$ and $\hat\gamma$, this effect has been used in the flat case analysed in ref.~\cite{Filev:2009xp} to show that the Goldstone modes of the theory obey a non-relativistic linear dispersion relation and to verify the modified counting rule valid for such modes. However, in the present work we will focus on the properties of the theory near the quantum critical point and in particular on the spectrum of fluctuations along $\chi$, since the spectrum of this mode contains a massless particle at the critical point responsible for the divergent correlation length. To this end we have to write the explicit expressions for the second order action derived from equations (\ref{S2DBI}), (\ref{S2WZ}) and derive the corresponding equations of motion.

\subsection{Quadratic Lagrangian}

Our results for the second order Lagrangian derived from equations (\ref{S2DBI}), (\ref{S2WZ}) are:
\begin{eqnarray}\label{L-F-F}
{\cal L}_{FF}&\propto&-\frac{\sqrt{-|E_0|}}{4}\,S^{\mu\mu'}S^{\nu\nu'}F_{\mu\nu}F_{\mu'\nu'}\ ,~~{\cal L}_{\beta\beta}\propto-\frac{\sqrt{-|E_0|}}{2}\,G_{\beta\beta}\,S^{\mu\nu}\partial_{\mu}\,\delta\beta\,\partial_{\nu}\delta\beta\ ,\\
{\cal L}_{\hat\alpha\hat\alpha}&\propto&-\frac{\sqrt{-|E_0|}}{2}\,G_{\hat\alpha\hat\alpha}\,S^{\mu\nu}\partial_{\mu}\,\delta\hat\alpha\,\partial_{\nu}\delta\hat\alpha\ ,~~{\cal L}_{\hat\gamma\hat\gamma}\propto-\frac{\sqrt{-|E_0|}}{2}\,G_{\hat\gamma\hat\gamma}\,S^{\mu\nu}\partial_{\mu}\,\delta\hat\gamma\,\partial_{\nu}\delta\hat\gamma \\\label{L-chi-chi}
{\cal L}_{\chi\chi}&\propto&-\frac{\sqrt{-|E_0|}}{2}\,\frac{G_{\chi\chi}\,G_{rr}}{G_{rr}+G_{\chi\chi}\,\chi_0'(r)^2}\,S^{\mu\nu}\,\partial_{\mu}\,\delta\chi\,\partial_{\nu}\delta\chi-\frac{1}{2}\frac{\partial\sqrt{-|E_0|}}{\partial\chi(r)^2}\,\delta\chi^2+\nonumber\\
&+&\frac{\sqrt{-|E_0|}}{2}\,\frac{G_{\chi\chi}}{G_{rr}}\,\chi_0'(r)^2\frac{\partial^2\log\sqrt{-|E_0|}}{\partial\chi_0(r)\,\partial\chi_0'(r)}\,\partial_r\delta\chi\,\delta\chi\ ,\\
{\cal L_{\hat\alpha\hat\gamma}}&\propto&-4\,H\,R^6\,\sin\alpha\sin\tilde\alpha\,\chi_0(r)^2\,\sqrt{1-\chi_0(r)^2}\,\partial_t\delta\hat\alpha\,\delta\gamma\ ,\\ \label{L-chi-F}
{\cal L}_{\chi F}&\propto& J^{\alpha\gamma}\frac{\partial \sqrt{-|E_0|}}{\partial\chi_0'(r)}\left(\partial_r\delta\chi\,F_{\alpha\gamma}-\partial_{\gamma}\delta\chi\,F_{\alpha r}+\partial_{\alpha}\delta\chi\,F_{\gamma r}\right)+J^{\alpha\gamma}\frac{\partial \sqrt{-|E_0|}}{\partial\chi_0(r)}\,\delta\chi\,F_{\alpha\gamma}=\nonumber \\
&=&-\partial_rJ^{\alpha\gamma}\frac{\partial \sqrt{-|E_0|}}{\partial\chi_0'(r)}\,\delta\chi\,F_{\alpha\gamma}+(\text{total derivative})\ ,
\end{eqnarray}
where $G_{mn}$ are the components of the then dimensional metric and the matrices $S\equiv(E_0^{-1}+{E_0^{-1}}^{T})/2$ and $J\equiv(E_0^{-1}-{E_0^{-1}}^{T})/2$ are given by:
\begin{eqnarray}\nonumber
&&S=\text{diag}\Big\{G_{tt}^{-1}\,,\,\frac{G_{\gamma\gamma}}{G_{\alpha\alpha}G_{\gamma\gamma}+H^2R^4\sin^2\alpha}\,,\,\frac{G_{\alpha\alpha}}{G_{\alpha\alpha}G_{\gamma\gamma}+H^2R^4\sin^2\alpha}\,,\,\frac{1}{G_{rr}+G_{\chi\chi}\chi_0'(r)^2}\,, \\
&&G_{\tilde\alpha\tilde\alpha}^{-1}\,,\,G_{\tilde\gamma\tilde\gamma}^{-1}\Big\}\ ,\\
&&J^{\mu\nu}=\frac{1}{\sin\alpha}\,\frac{H\,R^2}{r^4+H^2R^4}\left(\delta^{\mu}_{\gamma}\delta^{\nu}_{\alpha}-\delta^{\nu}_{\gamma}\delta^{\mu}_{\alpha}\right)\ .
\end{eqnarray}
Note that in deriving the last expression in equation (\ref{L-chi-F}) we have used the equation of motion for $\chi_0(r)$ and the Bianchi identity for the gauge field ($dF=0$). In the next subsection we use equations (\ref{L-F-F}),(\ref{L-chi-chi}) and (\ref{L-chi-F}) to show that for modes with vanishing angular momentum on the two sphere the equations of motion for $\delta\chi$ and the gauge field decouple. Next we obtain the spectrum of the fluctuations along $\chi$.

\subsection{Spectrum along $\chi$}
In this subsection we analyse the spectrum of the fluctuations along $\chi$. The reason to focus on this mode is that it contain the massless mode corresponding to the divergent correlation length at the critical point. As one can see from equation (\ref{L-chi-F}), the fluctuations along $\chi$ couple to the fluctuations of the gauge field. To obtain the corresponding equations of motion we vary the combined Lagrangian ${\cal L}_{\chi\chi}+{\cal L}_{\chi F}+{\cal L}_{F F}$. We observe that if one restricts the fluctuations to the ``S-mode'' by suppressing the angular momentum of the two sphere, the equations of motion for the fluctuations along $\chi$ decouple from those of the gauge field. To find the bound states we consider a consistent ansatz:
\begin{equation}
\delta\chi (t,r)=A\,\cos(\omega\, t)\,h_{\chi}(r)\ ,
\end{equation}
and solve numerically the resulting second order equation for $h_{\chi}(r)$. One can show that the general solution for $h_{\chi}(r)$ at infinity behaves as:
\begin{equation}\label{deltaChi-inf}
h_{\chi}(r)=\frac{a}{r}+\frac{b}{r^2}+\dots\ .
\end{equation}
Furthermore, substituting the asymptotic solution in the ``on-shell'' quadratic action one can show that only modes with $a=0$ are renormalisable and correspond to bound states with finite energy -- this is our quantisation condition ($a=0$). One can also show that the regular solution for $h_{\chi}(r)$ has predetermined boundary conditions for the first derivative at $r_0$ (the point where $\chi=1$ and the D5-brane closes). Therefore we are left with just one parameter specifying the solution, which corresponds to an overall amplitude (the equation is linear) which we set to one. Finally, we define the dimensionless parameters:
\begin{equation}
\tilde r =r/R;~~~\tilde \omega =\omega R\ .
\end{equation}
We are now ready to generate the numerical solution for $h_{\chi}(\tilde r)$.  For each value of the starting parameter $\tilde r_0$ we select those modes for which the coefficient $a$ in equation (\ref{deltaChi-inf}) vanishes. This gives us a discrete set of values for $\tilde \omega$ corresponding to the meson spectrum associated to the fluctuations along $\chi$. As one can expect, the spectrum depends crucially on the magnitude of the magnetic field $H$. Plots of the spectrum of the first three states as a function of the bare mass parameter (at fixed $H$) are presented in figures \ref{fig:fig13}, \ref{fig:fig14}, \ref{fig:fig15}. 
\begin{figure}[t] 
   \centering
\hspace{-0.7em}   \includegraphics[width=2.7in]{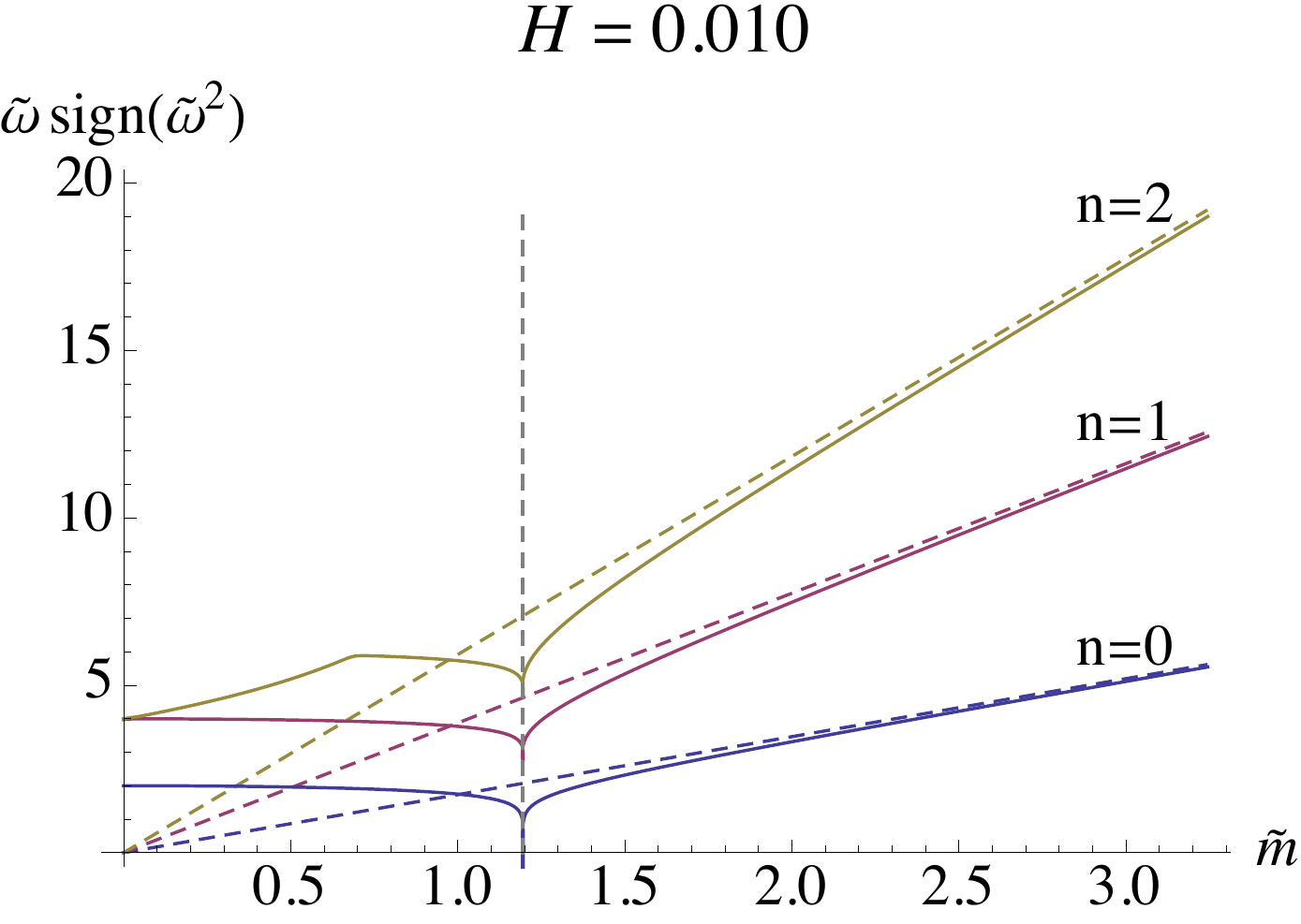} 
\hspace{1.4em}   \includegraphics[width=2.7in]{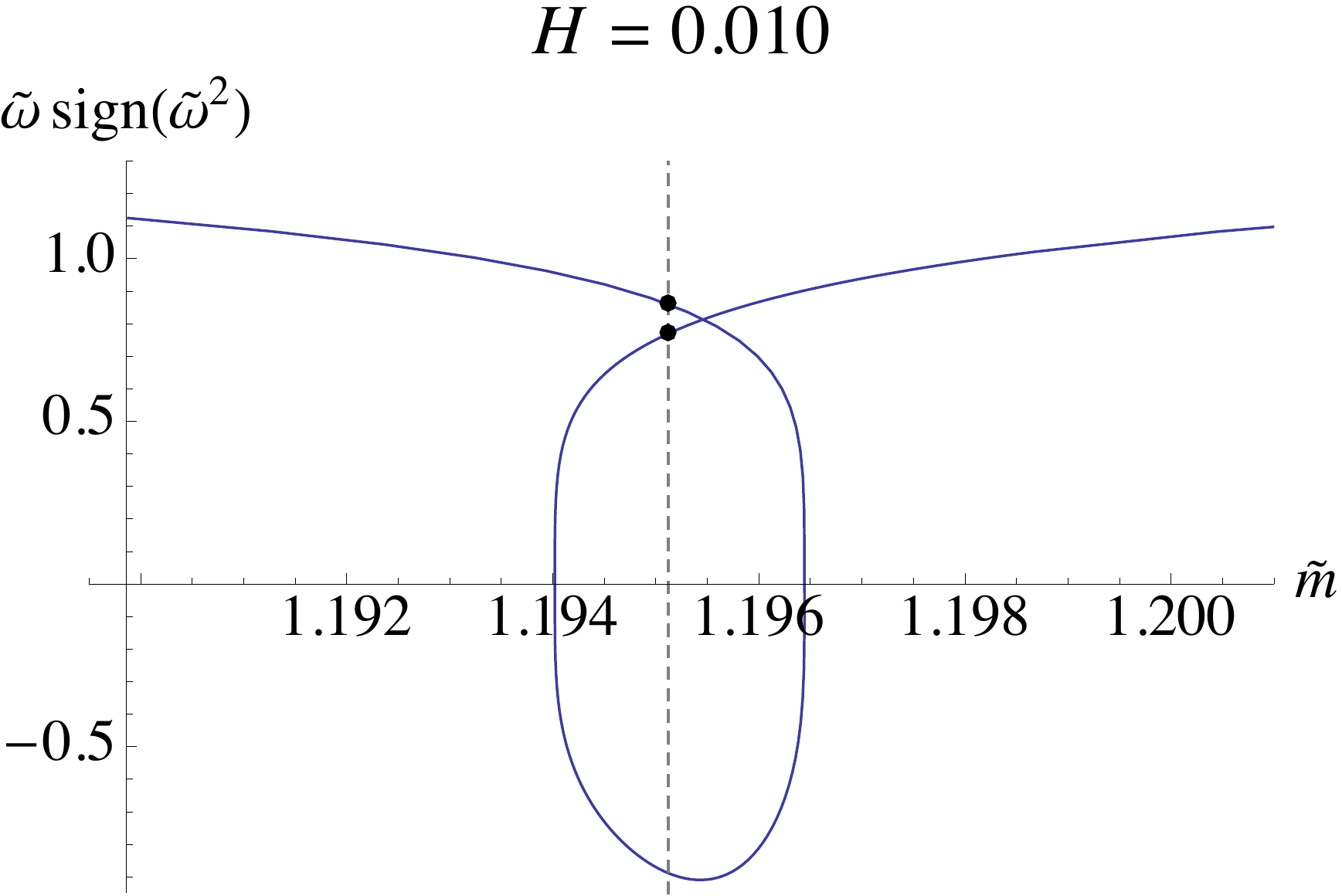} 
   \caption{\small Plot of the spectrum for $H=0.010$. The plots are for ${\rm Re}(\tilde\omega) \geq 0$ and $\tilde\omega^2\in {\rm Re}$. For large $\tilde m$ the spectrum approaches the flat result \ref{flat-o} represented by the dashed lines. The second plot represents the ground state zoomed in near the phase transition. The spectrum has a finite jump (between the small dots) and is tachyonic in the thermodynamically unstable regime.}
      \label{fig:fig13}
      \vspace{-.5em}
\end{figure}
One can see that for large bare mass parameter $\tilde m$ (and at vanishing magnetic field) the spectrum approaches the flat result first obtained in refs. \cite{Arean:2006pk}, \cite{Myers:2006qr}:
\begin{equation}\label{flat-o}
\omega =\frac{2\,m}{R^2}\sqrt{(n+1/2)(n+3/2)} ~~~~n=0,1,2,\dots\ ,
\end{equation}
which is represented by dashed lines in the figures. Another interesting feature of the spectrum is that at $\tilde m=0$ the ground state depends very weakly on the magnetic field and for the examples that we have studied ($H <0.10$) its value remains practically unchanged from the zero magnetic field result $\omega =2$, obtained in \cite{Erdmenger:2010zm}. However, the equidistant structure of the spectrum (at $\tilde m=0$, and $H=0$) is violated and the third state already deviates significantly. For higher excited states (or at stronger magnetic fields) we expect that a level crossing is taking place. 

The most interesting for us feature of the spectrum is its behaviour near the phase transition. In the second plot in figure \ref{fig:fig13} we have presented the spectrum of the ground state (for $H=0.010$) zoomed in near the phase transition. One can see that for the stable branches the spectrum has a discrete jump (between the black points in the figure), which is expected for a first order phase transition. Furthermore, one can see that the thermodynamically unstable phase of the corresponding Maxwell construction (see the shaded region in figure \ref{fig:fig6}) is tachyonic. There are also metastable regions close to the spinodal points on each side of the phase transition.  

\begin{figure}[t] 
   \centering
\hspace{-0.7em}   \includegraphics[width=2.7in]{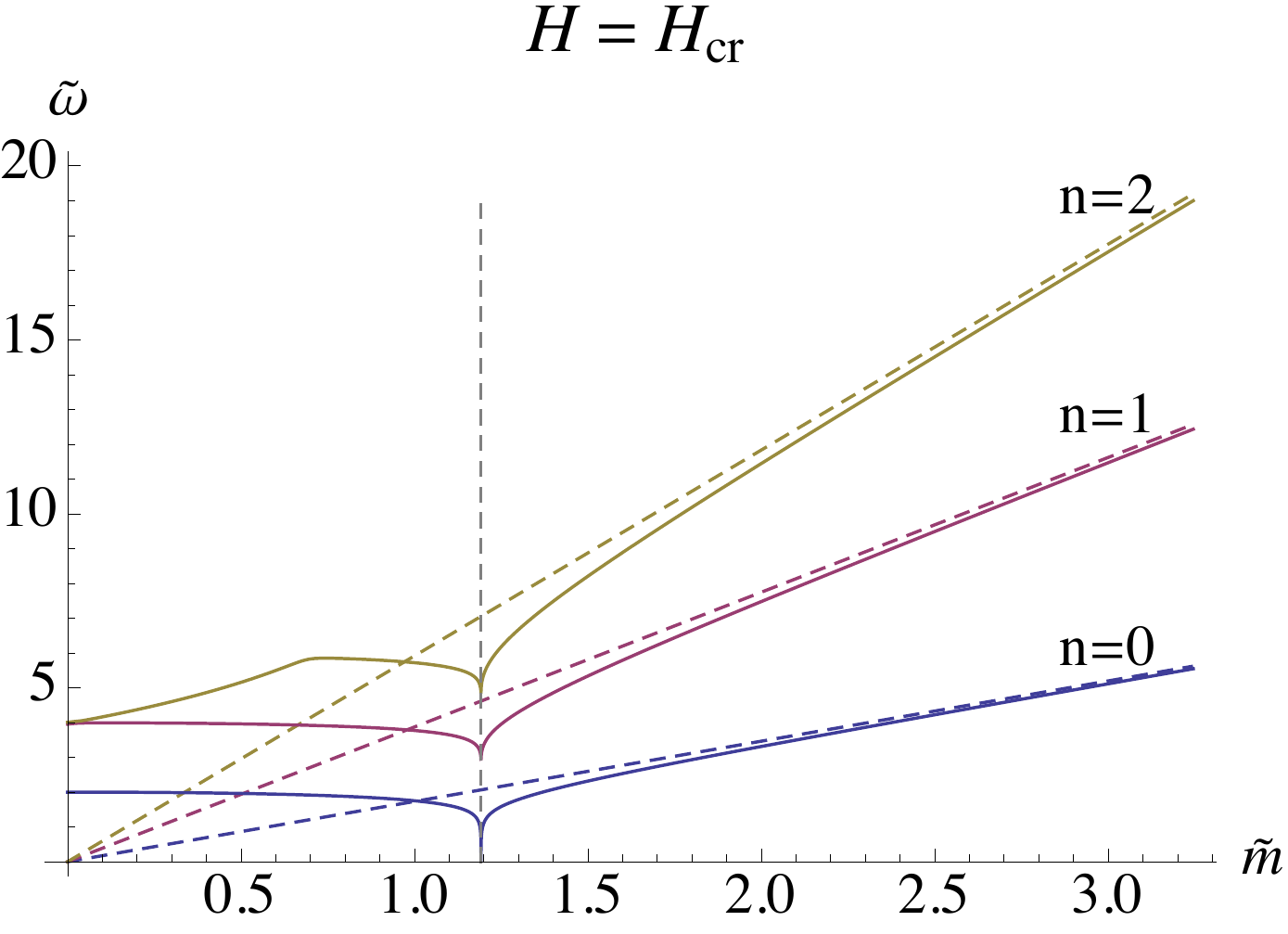} 
\hspace{1.4em}   \includegraphics[width=2.7in]{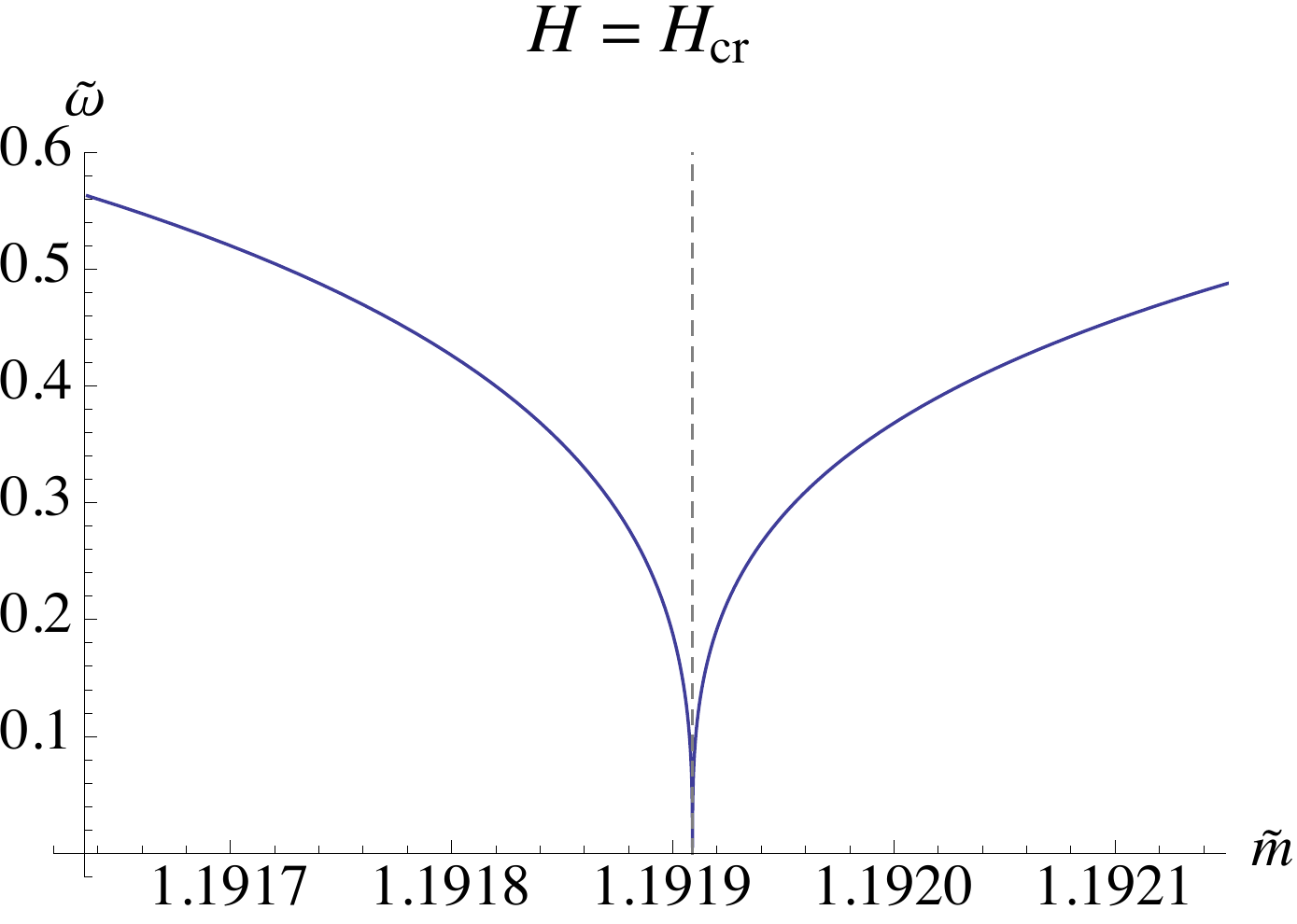} 
   \caption{\small Plot of the spectrum for $H=H_{cr}$. For large $\tilde m$ the spectrum approaches the flat result \ref{flat-o} represented by the dashed lines. The second plot represents the ground state zoomed in near the phase transition. The spectrum has a kink at the phase transition and vanishes. This massless mode corresponds to the diverging correlation length.}
      \label{fig:fig14}
\vspace{-.5em}
\end{figure}

In the second plot in in figure \ref{fig:fig14} we have presented the spectrum of the ground state for $H=H_{cr}$ near the second order phase transition. One can see that at the phase transition the spectrum has a non-analyticity and vansihes. Therefore, there is a massless mode in the spectrum at the critical point, which we associate to the divergent correlation length of the quantum fluctuations. We leave the study of the the corresponding critical exponent for the next subsection.

Finally, the second plot in figure \ref{fig:fig15} represents the ground state of the spectrum for $H=0.10$ (above the critical value $H_{cr}\approx 0.015769$) zoomed in near the region of the cross~over. As one can see the spectrum remains smooth at the cross~over. 

\begin{figure}[t] 
   \centering
\hspace{-0.7em}  \includegraphics[width=2.7in]{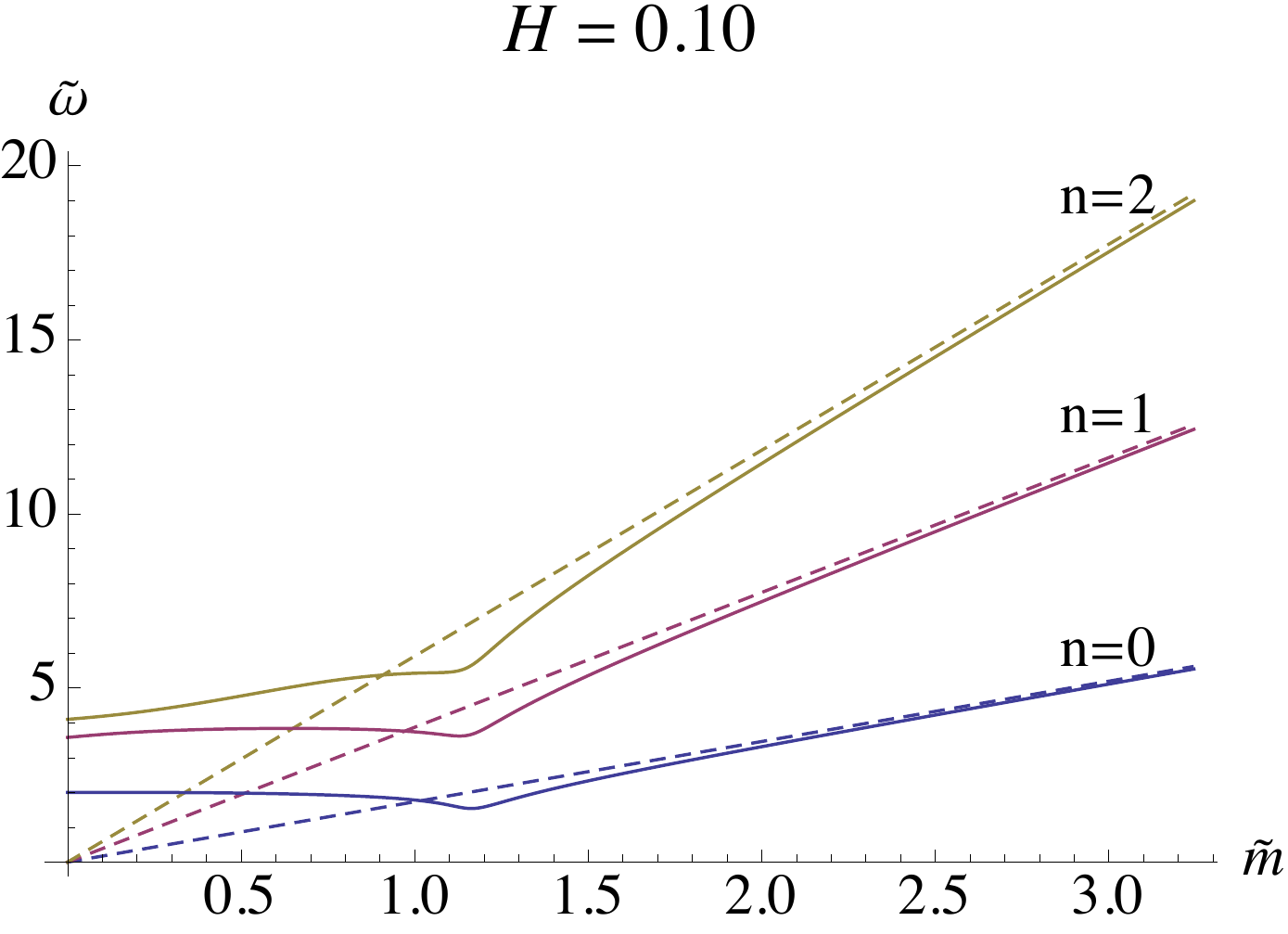} 
\hspace{1.4em}   \includegraphics[width=2.7in]{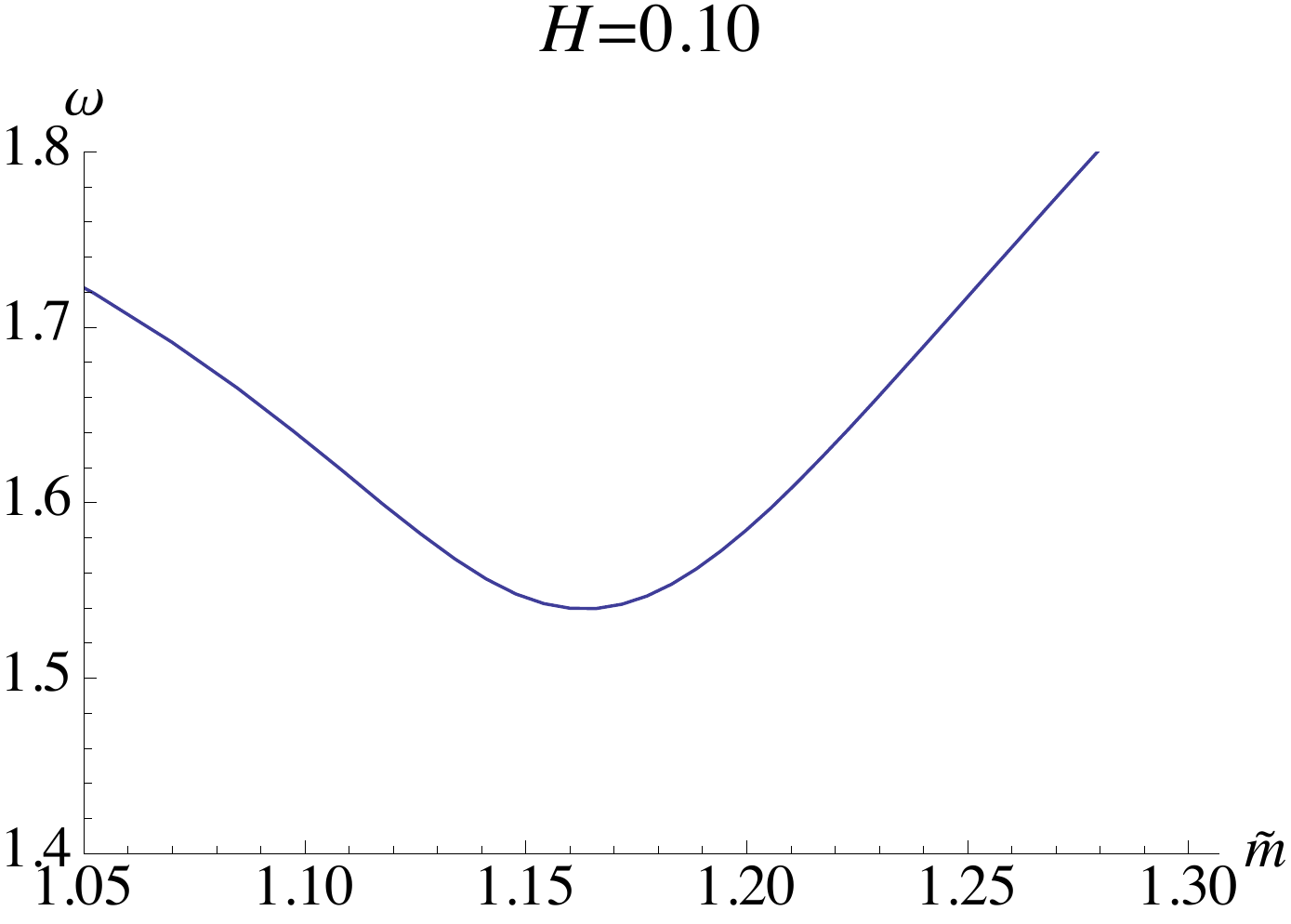} 
   \caption{\small Plot of the spectrum for $H=H_{cr}$. For large $\tilde m$ the spectrum approaches the flat result \ref{flat-o} represented by the dashed lines. The second plot represents the ground state zoomed in near the cross over. The spectrum remains smooth at the cross over.}
      \label{fig:fig15}
\end{figure}

\subsection{Critical behaviour of the spectrum}
In this subsection we analyse the critical behaviour of the spectrum at the phase transition. In particular we focus on the behaviour of the ground state near the phase transition. In figure \ref{fig:fig16} we have presented plots of $\tilde\omega$ vs $\tilde m-\tilde m_{cr}$ for some small internal $\Delta\tilde m$ near the critical parameter $\tilde m_{cr}$.
\begin{figure}[t] 
   \centering
\hspace{-0.7em}   \includegraphics[width=2.7in]{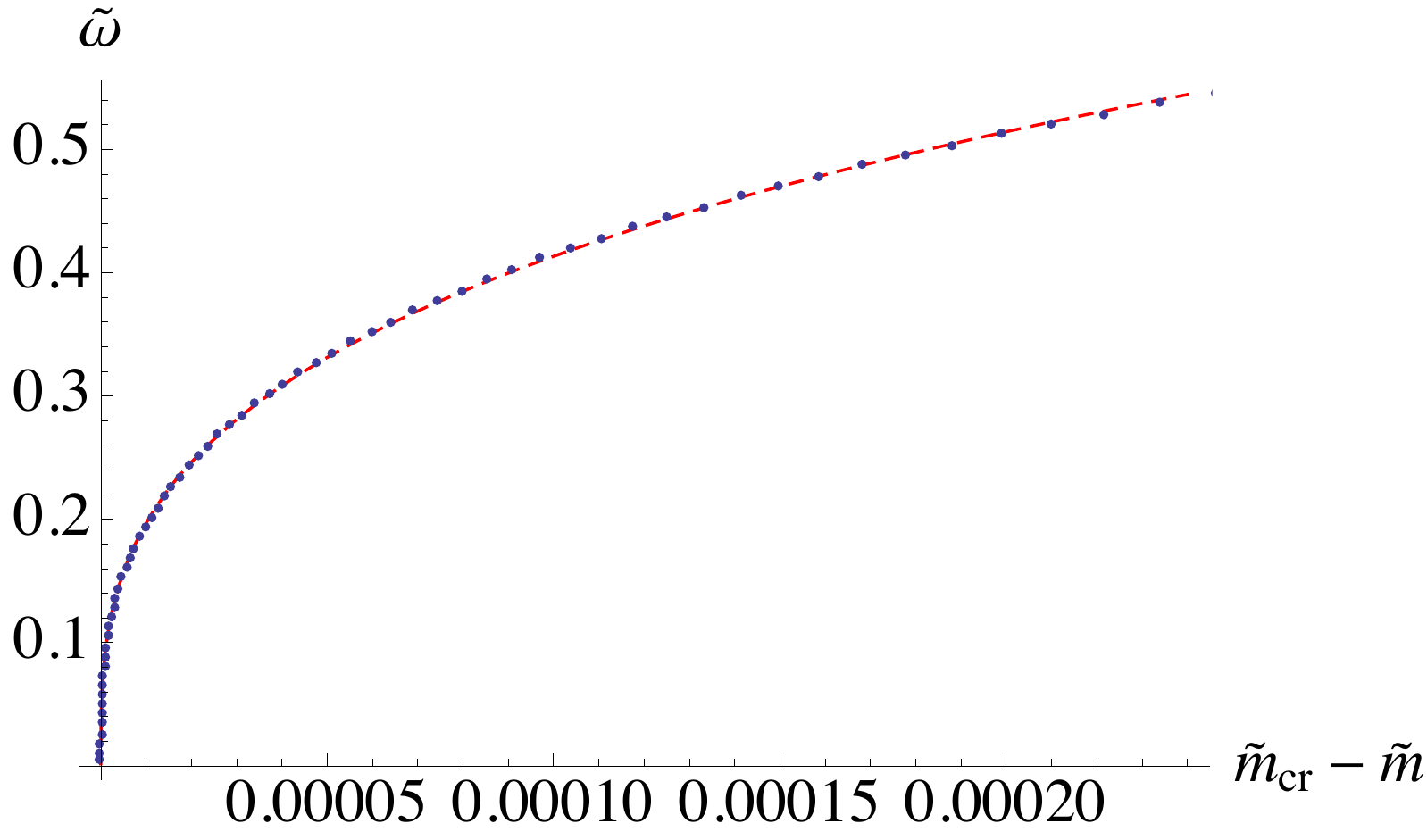} 
\hspace{1.4em}   \includegraphics[width=2.7in]{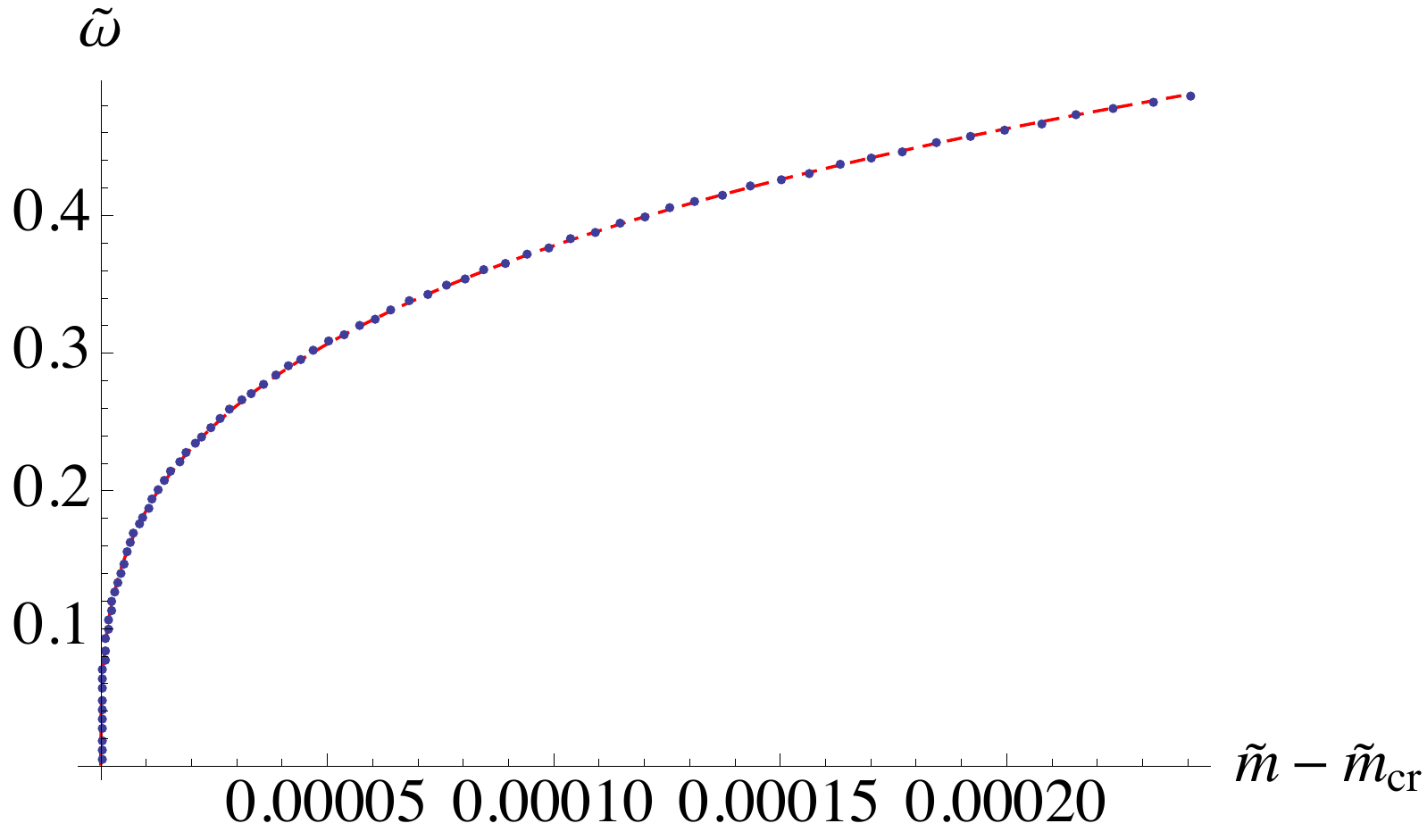} 
   \caption{\small Plots of the spectrum $\tilde\omega$ as a function of $\tilde m-\tilde m_{cr}$ on both sides of the phase transition. The dashed curves represent  ${\rm const_1} x^{1/3}+{\rm const_2} x^{2/3}$ fits.}
      \label{fig:fig16}
\end{figure}
The dashed curves represent ${\rm const_1} x^{1/3}+{\rm const_2} x^{2/3}$ fits, one can see the excellent agreement with the numerical data, suggesting that the spectrum approaches zero with a critical exponent $1/3$:
\begin{equation}
\omega\propto |\tilde m-\tilde m_{cr}|^{1/3}\ .
\end{equation}
 This also suggests that the parameter $\nu =\partial \omega/\partial m$ diverges at the phase transition as:
 \begin{equation}
\nu=\frac{\partial\omega}{\partial m}\propto\pm|\tilde m-\tilde m_{cr}|^{-2/3}\ .
 \end{equation}
This result, together with the results of section \ref{crit-exp}, complete our study of the critical exponents of the theory. It is intriguing that all diverging susceptabilities have critical exponent $-2/3$.

\section{Conclusion} \label{sec5}

In this paper we analysed the properties of a $2+1$ dimensional defect field theory on a two sphere, subjected to an external magnetic field. We found a rich phase structure and peculiar diamagnetic properties.

Our holographic set-up involved probe D5-branes in global AdS$_5\times S^5$ space-time. The external magnetic field was realised as a pure gauge $B$-field proportional to the volume form of the two sphere of the defect. For ``Ball'' embeddings the norm of the $B$-field diverges at the origin signalling the presence of a magnetic monopole. We showed that charge conservation of the RR flux requires that additional set of D3-branes be attached to the ``Ball'' embeddings, which is an unstable configuration. Remarkably, we found that the stable configuration is realised by Minkowski embeddings, which mimic the ``Ball'' embeddings and realise the D3-branes as a $\IR\times B_3\times S^2$ throat with small radius of the $S^2$, avoiding the divergence of the norm of the $B$-field.

Our numerical studies showed that for sufficiently small magnetic field, an analogue of the first order confinement/deconfinement phase transition at vanishing magnetic field is realised within the confined phase, represented by Minkowski embeddings. The phase transition does not correspond to a topology change transition of the D5-brane embeddings, allowing it to end on a critical point of a second order quantum phase transition, as the magnetic field is increased. Our system also has a non-zero negative condensate at vanishing bare mass, thus realising the effect of magnetic catalysis of chiral symmetry breaking. 

We showed that our set up realises a holographic quantum critical point and we found that all second derivatives of the free energy diverge with critical exponent of $-2/3$. Our study of the meson spectrum uncovered the existence of a massless mode at the phase transition signalling a divergent correlation length of the quantum fluctuations. The derivative of the meson mass with respect to the bare mass also diverges at the phase transition with a critical exponent of $-2/3$. 

Another intriguing property of our system is the existence of a persistent diamagnetic response, in the phase corresponding to small radius of the two sphere (small $\tilde m$). This behaviour bears resemblance to the properties of mesoscopic systems, such as nano tubes and quantum dots, which exhibit persistent diamagnetic current. 

One obvious direction for future studies is to complete the analysis of the meson spectrum of the theory. In particular it would be interesting to study the modified counting rule for goldstone bosons. In flat space the counting of the goldstone modes depends crucially on their dispersion relations \cite{Nielsen:1975hm}, it would be interesting to study these rules  for modes propagating on compact space. Since the flat case analogue of our system, realised goldstone modes with even dispersion relations \cite{Filev:2009xp} (see also refs.~\cite{Amado:2013xya} and~\cite{Amado:2013aea}), it is natural to expect that our system would describe their analogue on the two sphere.

Finally, it would be of a great interest to extend our studies to finite temperature. In particular it would be interesting to search for signatures of a quantum critical regime in the vicinity of the quantum critical point. The existence of such a regime is of a great experimental interest, since it is usually accompanied by exotic physical behaviour like novel non Fermi liquid phases. Given the full control on the system, that our holographic set up allows and the relative ease of introducing temperature (considering an AdS black hole), such studies can potentially have a large impact on our understandings of the physics of quantum critical points. 

\section*{Acknowledgments}

I would like to thank D. O'Connor and  P. Stamenov for useful comments and discussions and to A. O'Bannon for useful correspondence. Thanks are also due to D. Zoakos and M. Ihl for discussions and to D. Zoakos for reading the manuscript.






\end{document}